\documentclass[smallcondensed, final]{svjour3}                     

\RequirePackage{geometry}
\usepackage{amsmath}
\usepackage{multicol}
\usepackage{multirow}
\usepackage{tabularx}
\usepackage{float}  
\usepackage{subfloat}  
\usepackage[doublespacing]{setspace}  
\usepackage{graphicx, subfig}
\usepackage{caption}
\usepackage{dcolumn}
\usepackage{bm}
\usepackage{pdflscape}
\usepackage{booktabs}
\usepackage[figuresright]{rotating}

\usepackage[numbers]{natbib}

\newcommand{\beqa}{\begin{eqnarray}}
\newcommand{\eeqa}{\end{eqnarray}}
\newcommand{\beq}{\begin{equation}}
\newcommand{\eeq}{\end{equation}}
\newcommand{\be}{\begin{equation*}}
\newcommand{\ee}{\end{equation*}}
\newcommand{\bi}{\begin{itemize}}
\newcommand{\ei}{\end{itemize}}
\newcommand{\ben}{\begin{enumerate}}
\newcommand{\een}{\end{enumerate}}

\journalname{Flow, Turbulence and Combustion}

\begin{document}

\title{An LES study of turbulent flow over in-line tube-banks and comparison with experimental
  measurements}

\author{C.~Jin, ~I.~Potts,~D. C.~Swailes,~M. W.~Reeks}


\institute{C.~Jin, ~I.~Potts,~D. C.~Swailes,~M. W.~Reeks \at
School of Mechanical \& Systems Engineering, Newcastle University, Stephenson Building,
Claremont Road, Newcastle upon Tyne, NE1 7RU, UK\\
\email:{ian.potts@ncl.ac.uk}}

\titlerunning{An LES study and ... in-line tube-banks}

\date{Received: date / Accepted: date}

\maketitle

\begin{abstract}
  Turbulent flow across an in-line array of tube-banks with transverse
  and longitudinal pitch $P_T/D = 2.67$, and $~P_L/D = 2.31$,
  has been simulated successfully by Large
  Eddy Simulation (LES) based on the dynamic Smagorinsky
  subgrid scale model (SGS), in which a wall-layer model is
  used to reduce the computational cost. The flow
  structures across the tube-banks were examined through the normalized Q
  criterion. The surface pressure characteristics from the
  middle cylinder within each column of cylinders are found
  to agree well with the existing experimental data, as did also the
  values of drag and lift coefficients.  These
  results indicate that cylinders from the second column
  experience the minimum drag force and maximum lift force
  fluctuation. Spectral analyses were performed for velocity
  signals sampled behind each middle cylinder axis, which
  show that the dominant vortex shedding frequency does not
  vary across the tube-banks. On this basis, we
  also examined the shear layer instability. Finally, we
  report auto-correlation functions for streamwise and cross
  velocity fluctuations as a function of the spanwise length.
\end{abstract}


\section{Introduction}
Turbulent flow over tube-banks has been traditionally modelled using
the Reynolds-Averaged Navier-Stokes (RANS) equations with different
turbulence models
(see~\cite{beale1999numerical,rollet1999and,watterson1999predicting,benhamadouche2003coarse,wang2006turbulent}). The
flow across tube-banks is very unsteady, characterized by strong
vortex shedding and bluff-body
wakes. \citeauthor{rodi1997comparison}\cite{rodi1997comparison} has
demonstrated the difficulty or even impossibility of accurately
simulating the flow phenomena using the RANS methodology with the
standard $k-\epsilon$ turbulence model.  The astonishingly rapid
development of Large-Eddy Simulation (LES)
(see~\cite{rogallo1984numerical,mason1994large,sagaut2001large}) has
shown the potential to more accurately simulate simple flow phenomena
since LES resolves the large-scale unsteady motion directly and
requires only modelling of the universal small-scale turbulence
structures.

It is still a challenging task for numerical simulations to
obtain an accurate prediction of unsteady flow separation at
high Reynolds number across a single cylinder, not to mention
flow across tube-banks since the flow exhibits strong
unsteadiness and complex vortex structures. Whether the
dynamic boundary layer around a cylinder is accurately
resolved with appropriate numerical techniques and
resolution (see~\cite{davidson2009large, breuer1998large,
  lysenko2012large, lysenko2013large}) is
crucial to predicting the unstable region where turbulence is
generated, the instability of shear layer
(\citep{bloor1964transition}) and the physics of the wake
(\citep{williamson1996vortex, jordan2003resolving}). Hence the
conventional LES of turbulent flows across a single cylinder
and tube-banks is an extremely expensive endeavour at high
Reynolds number flow (\citep{breuer1998large,
  breuer2000challenging}). 

In recent years, LES has been used to simulate turbulent
flows across in-line and staggered
tube-banks (see~\cite{barsamian1997large, hassan2004tube,
  beale1999numerical, benhamadouche2003coarse,
  bouris1999two, liang2007large, lam2010experimental}) and
demonstrated its feasibility and effectiveness. \citeauthor{barsamian1997large}\cite{barsamian1997large} carried out a
two-dimensional LES calculation of flow over tube bundle
arrays using two subgrid scale models and studied the power
spectra and bound spectra of drag and lift forces. Later
extension in three-dimensional LES by \citeauthor{hassan2004tube}\cite{hassan2004tube}
was used to study velocity profile, power spectra density
(PSD) of velocities and forces, auto-correlation functions
of streamwise and transverse velocities in a flow past a
tube bundle at Reynolds number of $21700$ based on the free
stream velocity and cylinder diameter. \citeauthor{rollet1999and}\cite{rollet1999and}
performed LES based on a Finite Element Method for a
turbulent, incompressible flow around a staggered array of
tubes and compared the results with the measurements from \citeauthor{simonin1988measurements}
\cite{simonin1988measurements}. \citeauthor{beale1999numerical}\cite{beale1999numerical}
performed an LES of transient flow in a relatively low
Reynolds number regime of $Re \in \left[30, 3000\right]$
based on the gap velocity and cylinder diameter. Both
in-line square and staggered-square tube-banks were studied
in their work which accounted for pressure drop, lift, drag
and heat transfer. \citeauthor{liang2007large}\cite{liang2007large} employed an
unstructured grid Finite-Volume Method (FVM) based LES to
study the vortex shedding characteristics inside a staggered
tube bundle.

The simulation of turbulent flow over tube-banks can be
simplified to model merely a single circular cylinder,
provided that the cylinders are packed so closely that wake
vorticity does not turn up. In this case, the computational
domain is reduced to a single periodic circular cylinder
with four cylinder quarters around it. Hence, periodic
boundary conditions are assumed in the streamwise and
cross-flow direction. \citeauthor{benhamadouche2003coarse}\cite{benhamadouche2003coarse} carried
out a comprehensively comparative study of turbulent flow
across a single periodic cylinder in a tube bundle with LES,
coarse LES and URANS. In their study, LES with a wall
function modelling method gives the best results when
compared with
\citeauthor{simonin1988measurements}\cite{simonin1988measurements}'s
experimental data 
and DNS results from \citeauthor{moulinec2002diagonal}
\cite{moulinec2002diagonal}. \citeauthor{moulinec2004direct}\cite{moulinec2004direct}
carried out the diagonal Cartesian method (DCM) based DNS 
to study turbulent flow past an ``element cell'' in a tube-banks
with four sets of grids. The Reynolds number was equal to $6000$ based on the bulk velocity and the circular
diameter. They compared the results on the mean velocity and
r.m.s values from the finest cell with the datum measured by
\citeauthor{simonin1988measurements} \cite{simonin1988measurements} and numerical results calculated
by \citeauthor{rollet1999and}\cite{rollet1999and}, who has shown the feasibility of an
``element cell'' as an LES computational domain. Following
the work of \citeauthor{moulinec2004direct}\cite{moulinec2004direct},
\citeauthor{moulinec2004disappearing}\cite{moulinec2004disappearing} further investigated the wake
turbulence between a ``wide element'' consisting of $16$
circular cylinders using a three-dimensional DNS for $Re \in
\left[ 50, 6000\right]$ based on the bulk velocity.

In the present study, in contrast to previous research work
(see~\cite{rollet1999and,benhamadouche2003coarse,moulinec2002diagonal,
  moulinec2004disappearing}), a full scale turbulent flow
across an in-line tube-banks was computed with a
three-dimensional LES. The numerical technique was based on the
Finite-Volume Method (FVM) using wall-layer modelling on
unstructured grids with a collocated arrangement for all the
unknown flow variables. Particular attention was given to the
investigation of detailed statistics around the circular
cylinder in the middle cylinders of each column, which were
compared with the available experimental data of
\citeauthor{Shim1985} \cite{Shim1985},
\citeauthor{hill1986sources}\cite{hill1986sources} and \citeauthor{shim1988fluctuating}\cite{shim1988fluctuating}.

The rest of the this paper is structured as follows. The
computational methodology and geometry are presented first.
Then, a detailed comparison and discussion of mean and r.m.s
surface pressure distribution on the middle cylinders from
each column is given. In addition to that, the
corresponding drag and lift force, frequency analysis of
velocity signals and auto-correlations of streamwise and
cross-wise velocities in the spanwise direction, which
complement the existing experimental measurements, are
reported. Finally, conclusions are drawn.

\section{Computational methodology}

\subsection{Formulation of a dynamic Smagorinsky model}
The governing equations for LES are obtained by spatially
filtering the Navier-Stokes equations. In this process, the
eddies that are smaller than the filter size used in the
simulations are filtered out. Hence, the resulting filtered
equations govern the dynamics of large eddies in turbulent
flows. A spatially filtered variable that is denoted by an
overbar is defined using a convolution product
(see~\cite{leonard1974energy})
\beq
\overline{\phi}(\mathbf{x}, t) =
\underset{\mathcal{D}}\int %
\phi(\mathbf{y},t)G(\mathbf{x}, \mathbf{y}) d\mathbf{y}
\eeq
where $\mathcal{D}$ denotes the computational domain, and
$G$ the filter function that determines the
scale of the resolved eddies.

In the current study, the finite-volume discretization
employed itself provides the filtering operation as
\beq
\overline{\phi}(\mathbf{x}, t) = \frac{1}{\mathit{V}}\underset{\mathcal{D}}\int %
\phi(\mathbf{y}, t)d\mathbf{y}, \quad \mathbf{y} \in
\mathcal{V}
\label{filtering}
\eeq
where $\mathit{V}$ denotes the volume of a computational
cell. Hence, the implied filter function, $G(\mathbf{x},
\mathbf{y})$ in eq.\eqref{filtering}, is a top-hat filter
given by
\beq
G(\mathbf{x}, \mathbf{y}) = \begin{cases}
  1/\mathit{V} & \quad \text{for } |\mathbf{x} - \mathbf{y}| \in
  \mathcal{V} \\
  0 & \quad \text{otherwise}
  \end{cases}
\eeq
Filtering the continuity and Navier-Stokes  equations, the
governing equations for resolved scales in LES are obtained
\beq
\frac{\partial{\overline{u}}_i}{\partial{x}_i} = 0
\eeq
\beq
\frac{\partial{\overline{u}_i}}{\partial{t}} +
\frac{\partial{\overline{u}_i\overline{u}_j}}{\partial{x}_j}
= -\frac{1}{\rho}\frac{\partial{\overline{p}}}{\partial{x}_i} +
\frac{\partial}{\partial{x}_j}
\left ( \nu\frac{\partial{\overline{u}_i}}{\partial{x}_j}
\right ) - \frac{\partial{\tau_{ij}}}{\partial{x}_j}
\eeq
where $\tau_{ij}$ denotes the subgrid scale (SGS herefrom) stress tensor
defined by
\beq
\tau_{ij} = \overline{u_i u_j} - \overline{u}_i
\overline{u}_j
\label{sgs0}
\eeq
The filtered equations are unclosed since the SGS stress
tensor $\tau_{ij}$ is unknown.  The SGS stress tensor can
be  modelled based on
an isotropic eddy-viscosity model as:
\beq
\tau_{ij} -
\frac{1}{3}\tau_{kk} \delta_{ij} =
-2\nu_{t}\overline{S}_{ij}
\eeq
where  $\nu_t$ denotes the SGS eddy viscosity, and
$\overline{S}_{ij}$ is the resolved rate of strain tensor given
by
\beq
\overline{S}_{ij} = \frac{1}{2}\left (
  \frac{\partial{\overline{u}_i}}{\partial{x}_j} +
  \frac{\partial{\overline{u}_j}}{\partial{x}_i} \right )
\eeq
where $\nu_t$ is computed in terms of the \citeauthor{smagorinsky1963general}
\cite{smagorinsky1963general} type eddy-viscosity model
using
\beq
\nu_t = C_{\nu} \overline{\Delta}^2 |\overline{S}|
\eeq
where $C_\nu$ denotes the Smagorinsky coefficient, $|\overline{S}|$
 the modulus of rate of strain tensor for the resolved scales,
\beq
|\overline{S}| = \sqrt{2\overline{S}_{ij}\overline{S}_{ij}}
\eeq
and $\overline{\Delta}$ denotes the grid filter length obtained from
\beq
\overline{\Delta} = \mathit{V}^{1/3}
\eeq
Consequently, the SGS stress tensor is computated as
following
\beq
\tau_{ij} - \frac{1}{3}\delta_{ij} \tau_{ij} =
-2C_{\nu}\overline{\Delta}^2|\overline{S}|\overline{S}_{ij}
\label{smag1}
\eeq
This model claims to be simple and efficient.  It needs
merely a constant \textit{in priori} value for
$C_{\nu}$. Nevertheless, work from
\cite{lilly1966application,deardorff1970numerical,piomelli1988model}
has shown different values of $C_{\nu}$ for distinct
flows. Hence, the major drawback of the model used in LES is
that there is an inherent inability to represent a wide
range of turbulent flows with a single value of the model
coefficient $C_{\nu}$. Given that the turbulent flow over
tube-banks in the present study is fully three-dimensional,
the standard Smagorinsky SGS model is not used here to
compute the coefficient $C_{\nu}$.

\citeauthor{germano1991dynamic}\cite{germano1991dynamic} proposed a new procedure to
dynamically compute the model coefficient $C_{\nu}$ based
on the information obtained from the resolved large scales
of motion. The new procedure employes another coarser filter
$\widetilde{\Delta}$ (test filter)
whose width is greater than that of the default grid
filter. Applying the test filter to the filtered
Navier-Stokes equations, one obtains the following
equations
\beq
\frac{\partial{\widetilde{\overline{u}}_i}}{\partial{t}} +
\frac{\partial{\widetilde{\overline{u}}_i}\widetilde{\overline{u}}_j}{\partial{x}_j}
= -\frac{1}{\rho}\frac{\partial{\widetilde{\overline{p}}}}{\partial{x}_i} +
\frac{\partial}{\partial{x}_j}
\left ( \nu\frac{\partial{\widetilde{\overline{u}}_i}}{\partial{x}_j}
\right ) - \frac{\partial{\mathit{T}_{ij}}}{\partial{x}_j}
\eeq
where the tilde denotes the test-filtered
quantities. $\mathit{T}_{ij}$ represents the subgrid scale
stress tensor from the resolved large scales of motion and is given
by
\beq
\mathit{T}_{ij} = \widetilde {\overline{u_i u_j}} - \widetilde{\overline{u}}_i\widetilde{\overline{u}}_j
\label{sgs1}
\eeq
The quantities given in \eqref{sgs0} and \eqref{sgs1} are
related by the Germano identity:
\beq
\mathcal{L}_{ij} = \mathit{T_{ij}} - \widetilde{\tau}_{ij}
\label{germano}
\eeq
which represents the resolved turbulent stress tensor from
the SGS tensor between the test and grid
filters,$\mathit{T}_{ij}$ and $\tau_{ij}$. Applying the same
Smagorinsky model to $\mathit{T}_{ij}$ and $\tau_{ij}$, the
anisitropic parts of $\mathcal{L}_{ij}$ can be written
as
 \beq
 \mathcal{L}_{ij} - \frac{1}{3}\mathcal{L}_{kk}\delta_{ij} = -2CM_{ij}
 \eeq
 where
 \beq
 M_{ij} =
 \widetilde{\Delta}^2 |\widetilde{\overline{S}}|\widetilde{\overline{S}}_{ij}
 -
 \overline{\Delta}^2\widetilde{|\overline{S}|\overline{S}}_{ij}
 \label{germano1}
 \eeq
 One hence obtains the value of $C$ from \eqref{germano1}
 that is solved on the test filter level and then apply it to Eq.
 \eqref{smag1}.  The model value of  $C$ is
 obtained via a least squares approach proposed by
 \citeauthor{lilly1992proposed}\citep{lilly1992proposed}, since Eq. \eqref{germano1} is an
 overdetermined system of equations for the unknown variable $C$.  
\citeauthor{lilly1992proposed}\cite{lilly1992proposed} defined a criterion for minimizing
the square of the error as
\beq
 E = (L_{ij} - \frac{\delta_{ij}}{3}L_{kk} - 2CM_{ij})^2
\eeq
In order to obtain a local value, varying in time and space
in a fairly wide range, for the model constant $C$, one
takes $\frac{\partial{E}}{\partial{C}}$ and sets it zero to get
\beq
C = \frac{1}{2}\frac{L_{ij}M_{ij}}{M_{ij}M_{ij}}
\eeq
A negative C represents the transfer of flow energy from the
subgrid-scale eddies to the resolved eddies, which is known
as \textit{back-scatter} and
regarded as a desirable attribute of
the dynamic model.

\subsection{The Werner and Wengle wall layer model}
The Large Eddy Simulation (LES) of turbulent flow over tube-banks
is hampered by expensive computational cost incurred when the
dynamic and thin near-wall layer is fully resolved. To
obviate the computational cost associated with calculating
the wall shear stress from the laminar stress-strain relationship that
requires the first cell to be put within the range of $y^{+}
\approx 1$, \citeauthor{werner1993large}\cite{werner1993large} proposed a simple power-law to
replace the law of the wall, in which the velocity
profile on a solid wall is given as following,
\beq
u^{+} = \begin{cases}
  y^{+} & \quad \text{for } y^{+} \le 11.81\\
  A(y^+)^B & \quad \text{for } y^{+} > 11.81
\end{cases}
\label{ww-model}
\eeq
where $A = 8.3$ and $B=1/7$. An analytical integration of Eq.
\eqref{ww-model} results in the following relations for the
wall shear stress
\beq
|\tau_w| =
\begin{cases}
  \frac{2\mu|u_p|}{\Delta y} & \quad \text{for } y^{+} \le 11.81\\
  \rho \left [ \frac{1- B}{2}A^{\frac{1+B}{1-B}}\left
      (\frac{\mu}{\rho \Delta y}\right )^{1+B} 
     + \frac{1 + B}{A} \left( \frac{\mu}{\rho \Delta
      y}^B |u_p|\right )\right ]^{\frac{2}{1+B}}  & \quad \text{for } y^{+} > 11.81
\end{cases}
\label{ww-model}
\eeq
where $u_p$ is velocity component parallel to the wall and given by:
\beq
|u_p| = \frac{\mu}{2\rho\Delta y}A^{\frac{2}{1-B}}
\eeq

\subsection{Flow configuration of in-line tube-banks}
\begin{figure}[t!bp]
  \centering
  \includegraphics[trim=0.00in 6.0in 0.0in
  0.0in,clip=true,scale=1.0, angle = 0, width = 1.0\textwidth]{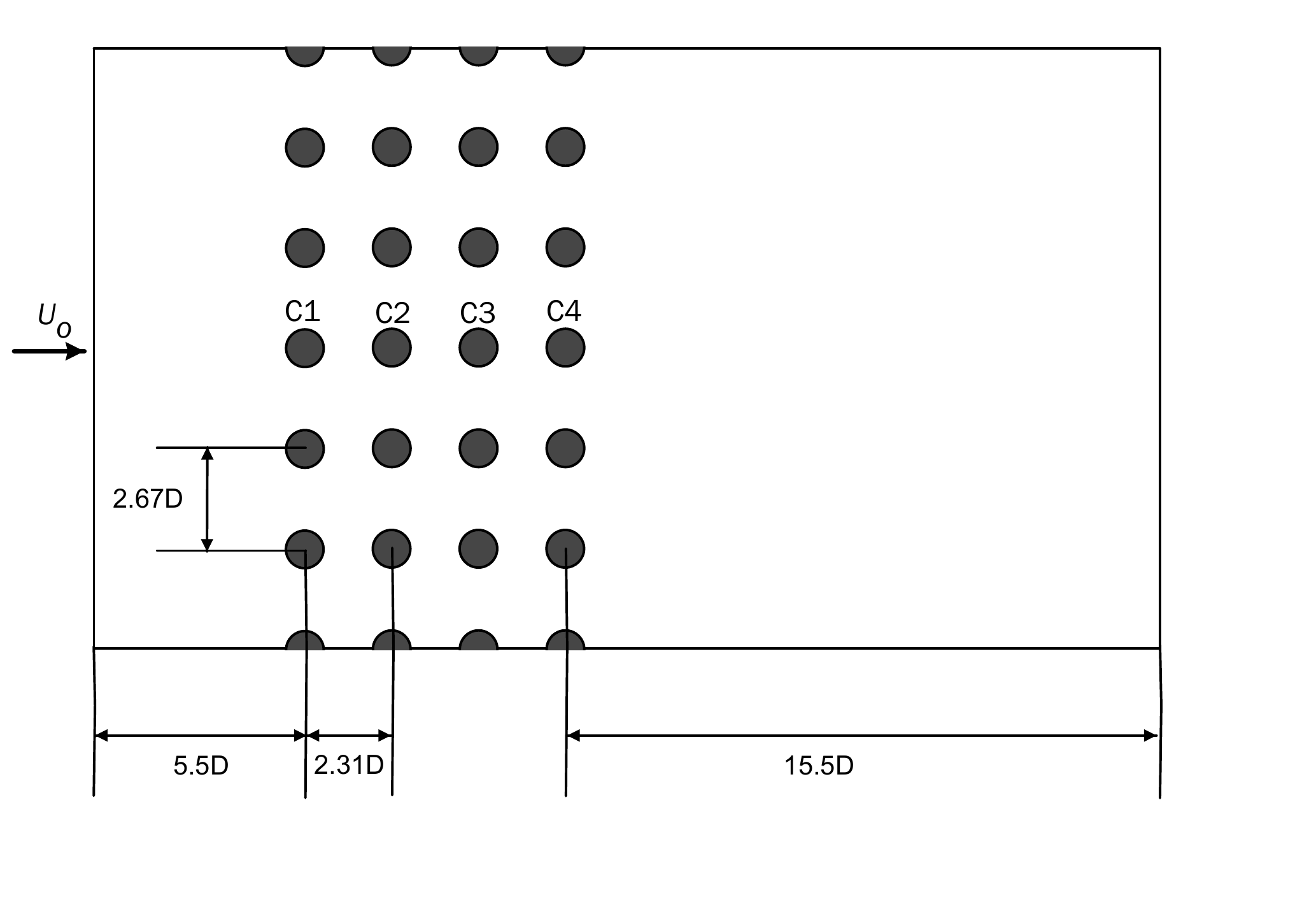} 
  \caption[Configuration of the four-column in-line
  tube banks]{Configuration of the four-column in-line
  tube banks}
  \label{fig:configuration}
\end{figure}

\begin{figure}[htb]
  \centering
  \includegraphics[trim=0.3in 0.6in 0.3in
  0.3in,clip=true,scale=0.6, angle = 0]{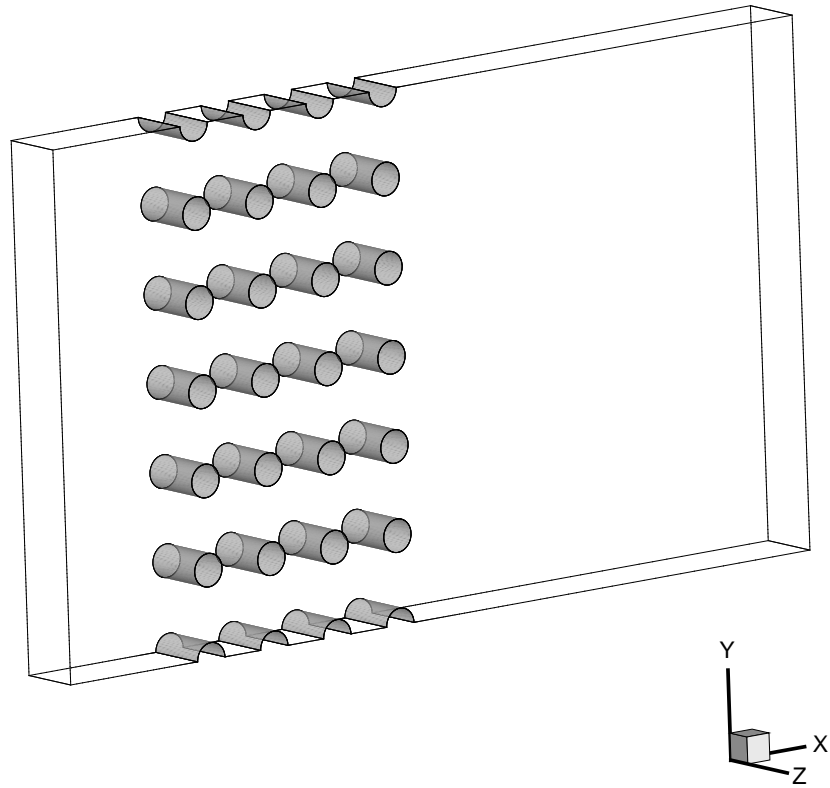} 
  \caption[Configuration of the four-row in-line
  tube bank, The $x-$ axis indicates the freestream flow
  direction; $y-$ and $z-$axis respectively indicate the
  transverse and spanwise direction.]{Configuration of the four-row in-line
  tube bank, The $x-$ axis indicates the freestream flow
  direction; $y-$ and $z-$axis respectively indicate the
  transverse and spanwise direction.}
  \label{fig:configuration1}
\end{figure}

The flow configuration is shown in
figure~\ref{fig:configuration} and the coordinate system
depicted in figure~\ref{fig:configuration1}. Flow is from
left to right and normal to the cylinder axis. The
computational domain is of size $L_x \times L_y \times L_z =
28D \times 16D \times 2D$, where $D$ denotes the cylinder
diameter. This configuration is based on the second test
case considered in \citeauthor{Shim1985}\cite{Shim1985} which measures surface
pressure distributions and fluctuating lift forces and was
performed in a suction-type wind tunnel. It consists of
four-column in-line tube bundles with transverse
pitch-to-diameter ratio ($P_T/D $) $S_T$ of 2.67 and
longitudinal pitch-to-diameter ratio ($P_L/D$) $S_L$ of 2.31,
respectively. The Reynolds number $Re_o$ based on the free
stream velocity $U_o$ and the cylinder diameter $D$ equals
to $9600$, and $Re_g$ based on the gap streamwise velocity
between two cylinders is equal to $15200$.

The Navier-Stokes solver used in this work uses a
cell-centered, collocated grid arrangement finite-volume
(FV) discretization method.  All spatial terms in the
momentum equations are discretized by the bounded central
differencing scheme, which not only boasts the advantage of
low numerical diffusion of central-differencing scheme but
also eliminates unphysical oscillations in the solution
fields. The spatial discretization scheme is
based on a multi-dimensional, least squares cell-based
gradient reconstruction scheme to guarantee a second-order
spatial accuracy. In order to prevent unphysical
checker-board pressure field, This study employs a procedure
similar to that proposed by \citeauthor{rhie1983numerical}\cite{rhie1983numerical}. The
Gear's implicit, three-level second-order accurate scheme is
employed for temporal discretization. A generalized
fractional-step method is employed for the overall
time-advancement.

\begin{sidewaysfigure}
\vspace{6.0in}
  \begin{center}
   \hspace{-1.0cm}
    \includegraphics[trim=0.8in 0.4in 0.0in
  0.0in,clip=true,scale=0.8, angle = 0]{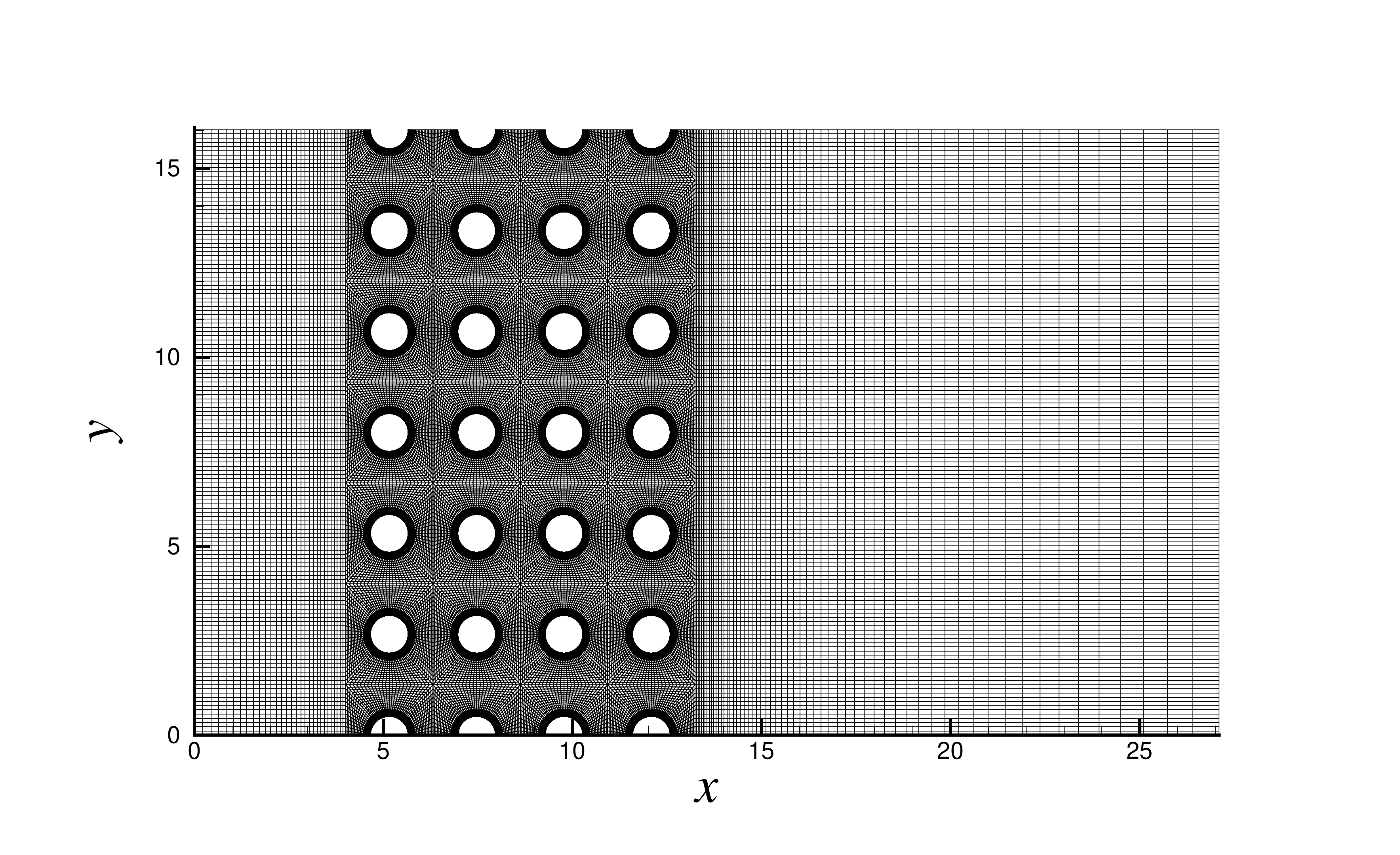} 
\end{center}
  \caption[Computational domain and mesh in the
    $x-y$ plane]{Computational domain and mesh in the
    $x-y$ plane}
  \label{fig:grid}
\end{sidewaysfigure}

The computational grid is evident in
figure~\ref{fig:grid}. The total number of grid elements
used for the present simulation is $2730240$. the mesh has
an embedded region of fine mesh designed for each cylinder
in order to enhance the mesh resolution near the cylinder
without incurring too large an increase in the total number
of mesh elements.  96 grid points hence are allocated along
the cylinder surface. The gird spacing on the cylinder in
the radial, circumferential, and spanwise direction are $\Delta r/D
= 1.4 \times 10^{-2}$, $\Delta \theta/D = 3.27 \times
10^{-2}$, $\Delta z/D = 5.0 \times 10^{-2}$,
respectively. The first cell adjacent to the cylinder is
within the range $\Delta y^{+} < 11.8$ in wall
units\footnote{The superscript $+$ denotes a non-dimensional
  quantity scaled using the wall variables, e.g. $y^{+} =
  yu_{\tau}/\nu$, where $\nu$ is the kinematic viscosity and
  $u_\tau = \sqrt{\tau_w/\rho}$ is the wall friction
  velocity based on the wall shear stress ${\tau}_w$, and
  which is a velocity scale representative of velocities close
  to a solid boundary.}  that satisfies the requirements of
the Wener-Wengle wall-layer model for LES. Prior to the
present simulation, with the standard Smagorinsky subgrid
scale model, a coarser grid simulation were carried out to
determine the resolution.

With fully developed turbulent flow,
periodic boundary conditions are
justified to use along the normal ($y$) and spanwise ($z$)
direction. For the inlet boundary condition, a simple
uniform velocity profile is assumed and the turbulent
intensity set to zero. Hence, the turbulence fluctuations at
the inlet was not accounted for temporally and
spatially. Nevertheless, a length $5D$ before the first column
bank is used to allow the development of turbulence. At the
exit boundary, the solution variables from the adjacent
interior cells are extrapolated to satisfy the mass conservation.

The simulation is advanced with a non-dimensional time step
$\Delta t U_o/D \approx 2 \times 10^{-3}$ that yields
maximum Courant-Friedrichs-Lewy (CFL) number of 0.5. For results
presented here, the first-order statistics are collected by
integrating the governing equations over an interval of
$30D/U_o$, and all the statistics are averaged over the $40$
sampling points along the spanwise direction.

\section{Results and discussions}

To provide an overview of the development of turbulent flow 
across the four-column in-line tube-banks, wake vortices
visualized using the Q criterion (see \cite{hunt1988eddies}) are
presented first.  Then, time-resolved pressure distributions
provide quantitative information on surface pressure
fluctuations, which are compared with experimental
measurements (\citep{Shim1985}). Following this, the
time histories of coefficient of drag $C_D$ and lift are
given. The development of vortex shedding behind the
cylinder in the middle column are investigated via examining the
corresponding energy spectrum in the wake. The coherence of
vortex shedding along the length of the middle cylinder is
studied through computing the auto-correlation function
of each velocity fluctuation component. In the present work,
the turbulent flow across tube-banks has been considered to
have reached the statistically stationary state after a
simulation time of $T = 200D/U_o$. All the statistics
presented here are computed after this transient
stage. Further, the statistics are averaged in the periodic
spanwise direction.

\subsection{Instantaneous flow field}
\begin{figure}[h!tbp]
  \centering
  \includegraphics[trim=0.6in 1.35in 0.00in
  0.00in,clip=true,scale=0.66, angle = -90]{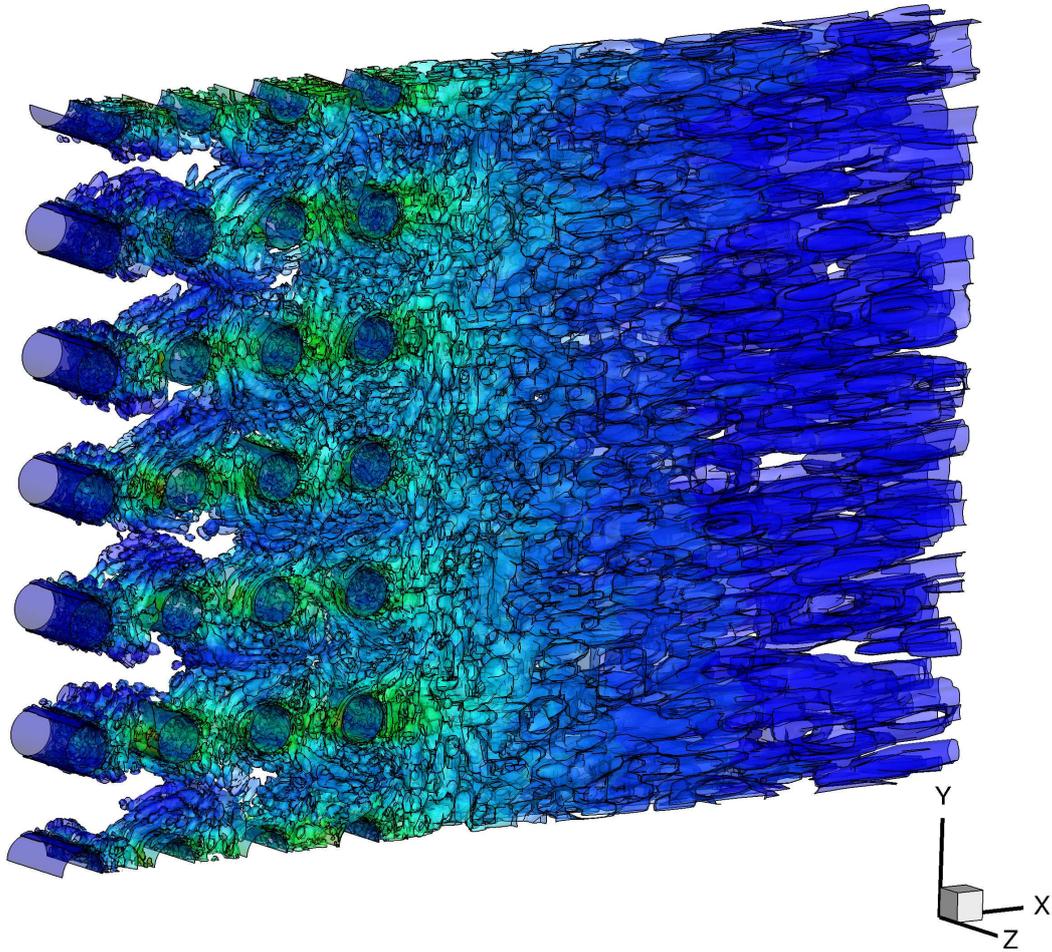} 
  \caption[The filtered flow structure development across the
    four-in-line tube banks, iso-surface of the second
    invariant of velocity gradient tensor, colored by the
    resolved turbulent kinetic energy (TKE)]{The filtered flow structure development across the
    four-in-line tube banks, iso-surface of the second
    invariant of velocity gradient tensor, colored by the
    resolved turbulent kinetic energy (TKE)}
  \label{fig:nq-tke}
\end{figure}

The contours of turbulent kinetic energy (TKE) at a given time across the
four-in-line tube banks are presented in
figure \ref{fig:nq-tke} using a $normalized~Q-criterion = 8
\times 10^{-2}$. The Q-criterion, proposed by
\citeauthor{hunt1988eddies}\cite{hunt1988eddies}, is defined as the second invariant of
velocity gradient tensor $\nabla \mathbf{u}$ for incompressible flows by the
following expression
\beq
Q =
\frac{\partial{u_i}}{\partial{x_j}}\frac{\partial{u_j}}{\partial{x_i}}
= \frac{1}{2}\omega_{i}^2 - e_{ij}^2
\eeq
where $e_{ij}^2$ and ${\omega_{i}}$ denotes the symmetric and
antisymmetric parts of $\nabla \mathbf{u}$, respectively.

The instantaneous flow field shows the salient feature of
the wake dynamics where a wide, yet different range of scales
behind every column of cylinders can be observed. As far as the
first column cylinders are concerned, the flow shows no
unexpected properties, but a few points are worth noting for
comparison with flow patterns behind other
cylinders. 

\begin{figure}[h!tbp]
  \centering
  \includegraphics[trim=1.6in 1.35in 0.5in
  0.5in,clip=true,scale=0.66, angle = -90, width = 1.0\textwidth]{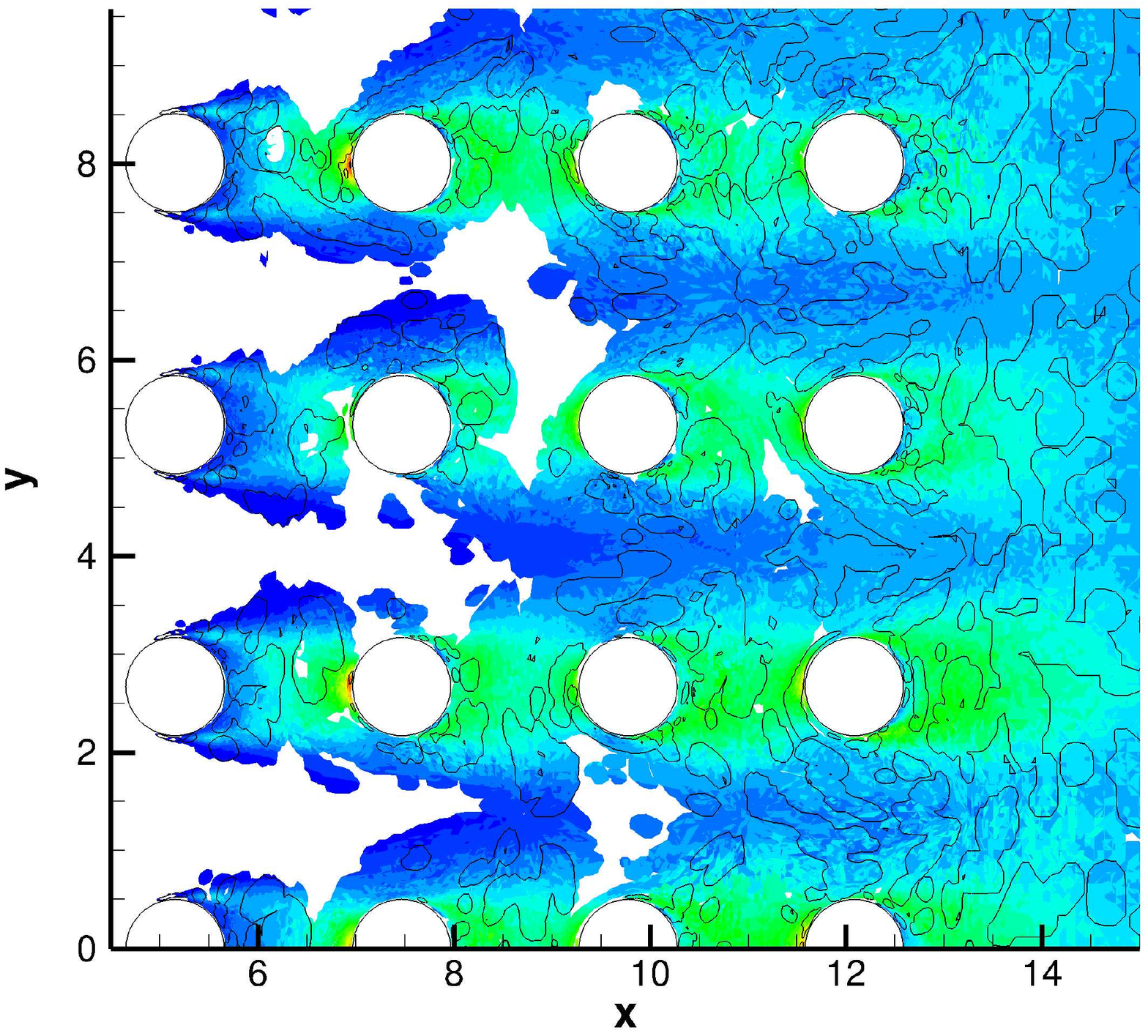} 
 \caption[Vortex motion around cylinders at
    the middle plane cut]{Vortex motion around cylinders at
    the middle plane cut}
  \label{fig:vortex_side_view}
\end{figure}

Firstly, the boundary layer on each individual
cylinders of the first column remains laminar up to the
separation point, and it undergoes transition to turbulence
in the separated shear layer. Whilst the boundary layer
separation on the cylinders from the downstream columns is much
delayed and so that the wake is much narrow, resulting in a
much smaller coefficient of drag. This principally results
from the inflow conditions for the downstream
cylinders. Figure \ref{fig:vortex_side_view} shows a close-up of the
vortex motion around the cylinders across the middle plane,
again shown contours of TKE in terms of the same
normalized-Q criterion.  It is evident that the turbulence
level is quite high at the front side of the downstream
cylinders.

Secondly, figure \ref{fig:nq-tke} illustrates different flow
pattern of vortex travelling downstream each column of
cylinders. Large coherent structures are visible in the wake
of first column of cylinders. Nevertheless, the classical von
Karman vortex streets fail to arise because the second
cylinder column lies within the range of the recirculation
region of flow behind the first column and hence suppresses the
vortex street formation in the wake. Another effect of the downstream
cylinder is to increase the wake instabilities further. Large
flow structures are lost and broken into small eddies,
producing ultimately a fully developed grid turbulence after
the final cylinder column.

\subsection{Surface pressue characteristics}
\label{sec:spd}
\begin{sidewaysfigure}   
\vspace{6.0in}
\centering
  \subfloat{\label{fig:Cp0}\includegraphics[trim=0.00in
    0.00in 0.00in 0.00in,clip=true,scale=1.0, width = 0.4\textwidth]{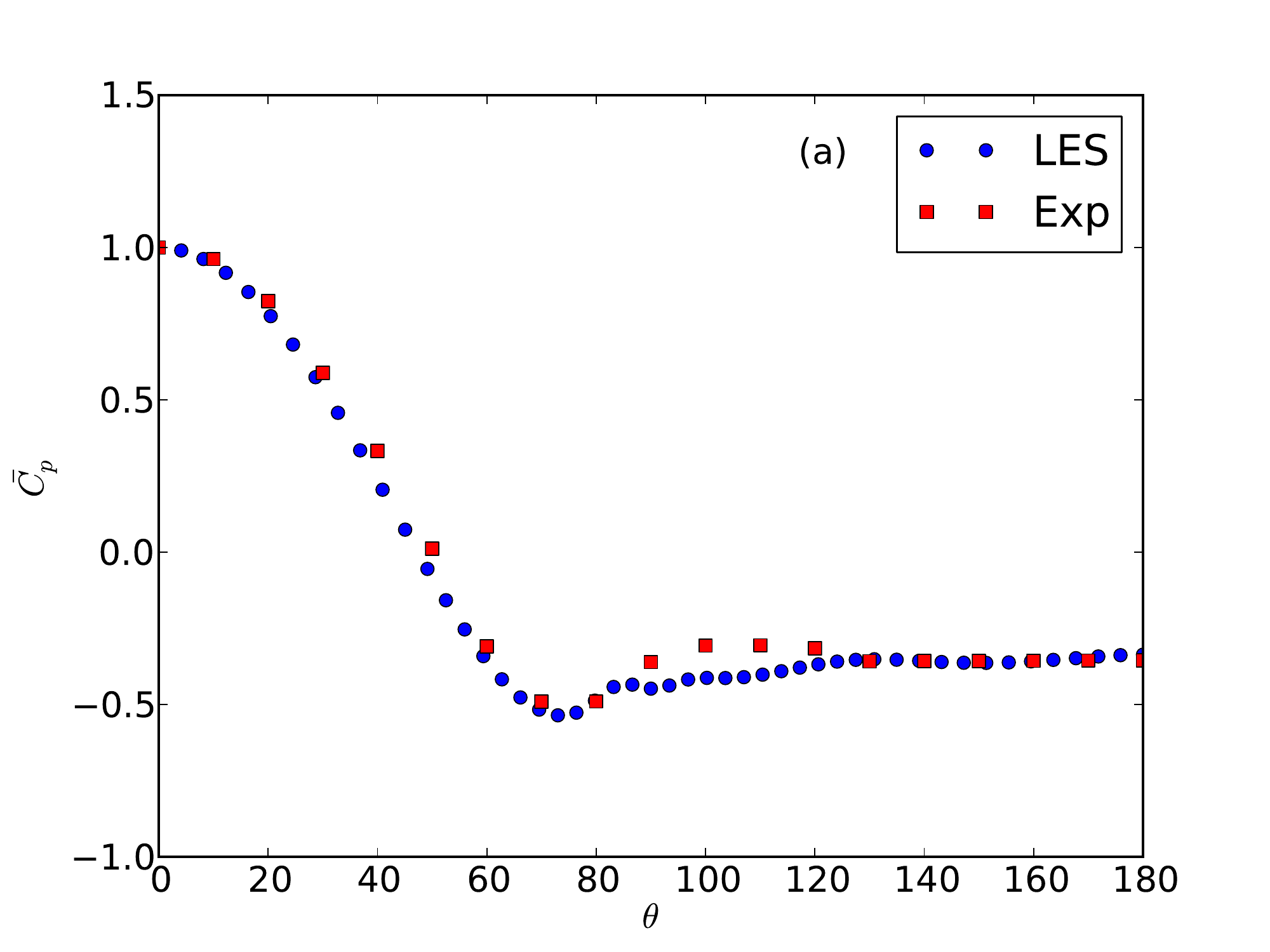}} 
  \subfloat{\label{fig:Cp1}\includegraphics[trim=0.00in
    0.00in 0.00in 0.00in,clip=true,scale=1.0, width =
    0.4\textwidth]{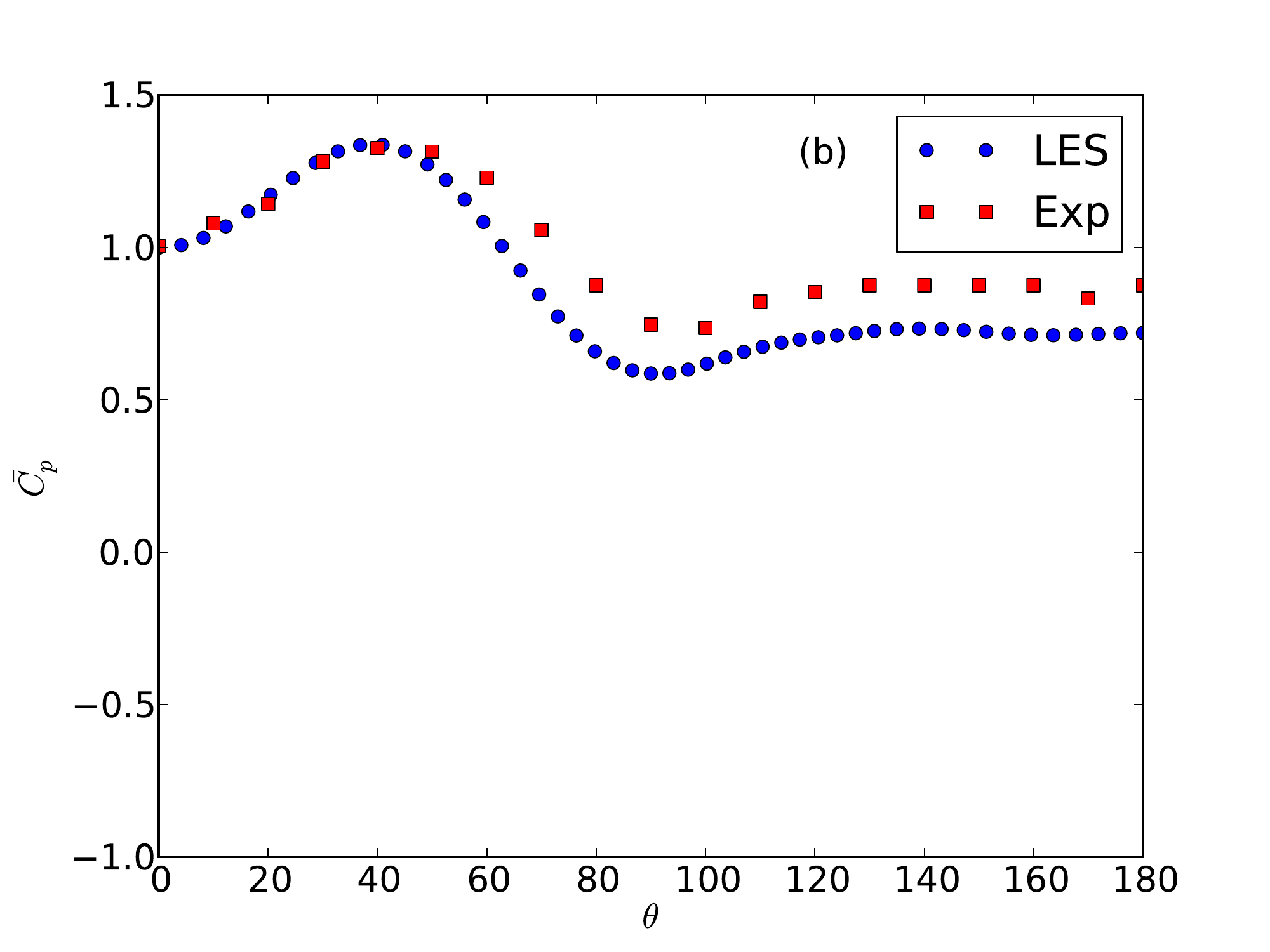}} \\
  \subfloat{\label{fig:Cp2}\includegraphics[trim=0.00in
    0.00in 0.00in 0.00in,clip=true,scale=1.0, width = 0.4\textwidth]{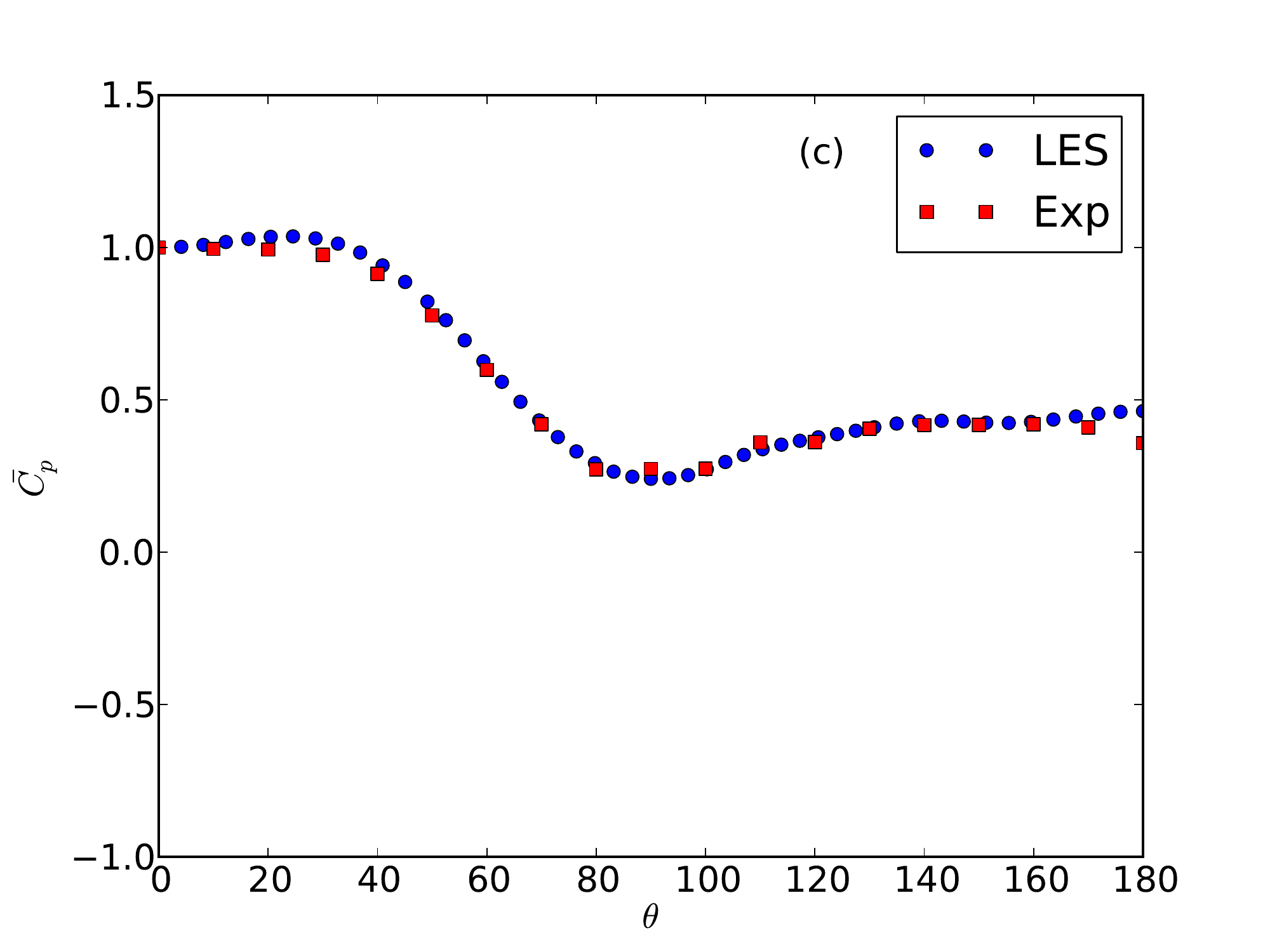}}  
  \subfloat{\label{fig:Cp3}\includegraphics[trim=0.00in
    0.00in 0.00in 0.00in,clip=true,scale=1.0, width = 0.4\textwidth]{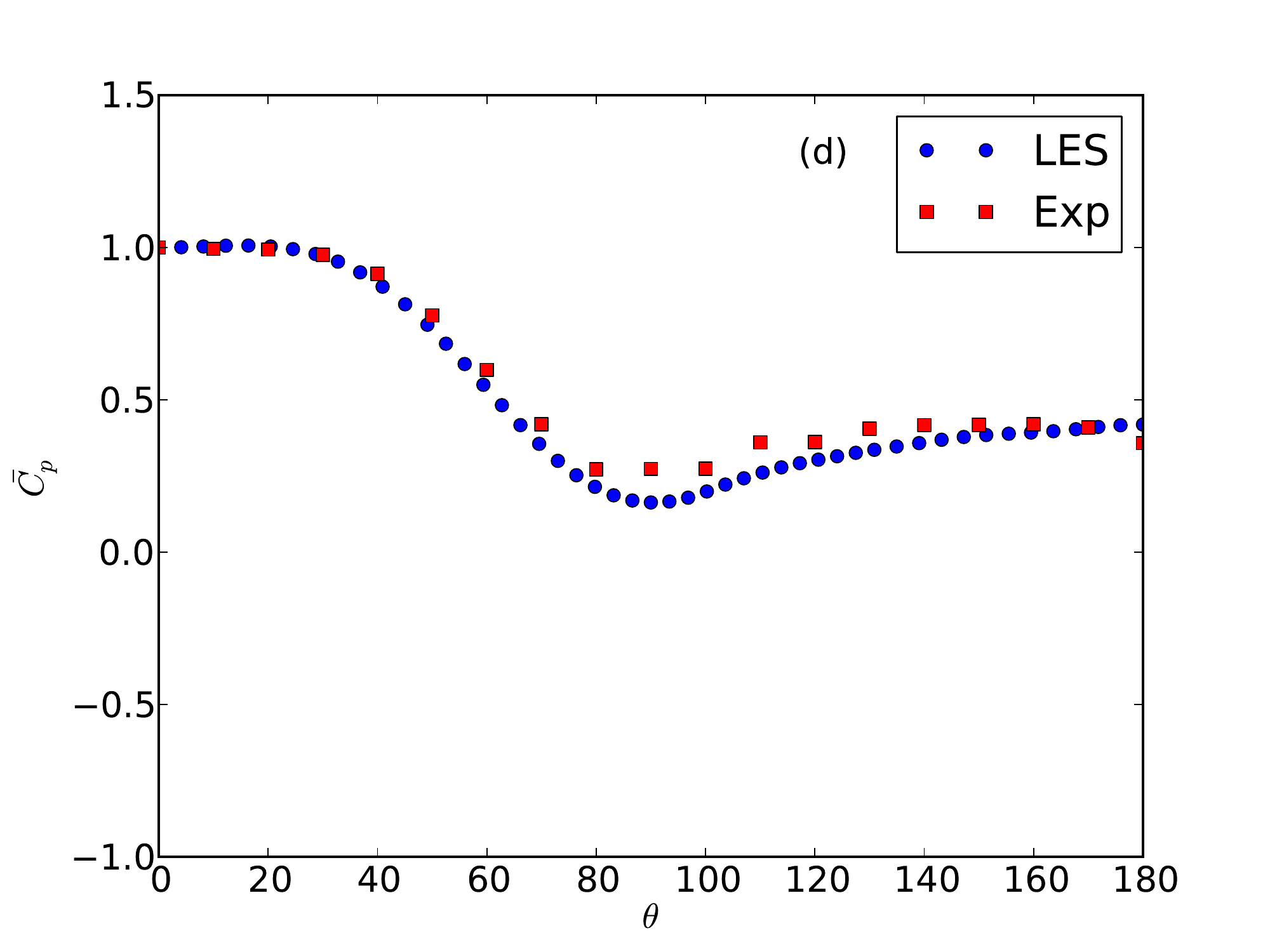}}

  \caption{Averaged mean $C_p$ as a function of angle from
    the front stagnation point, (a)~\textbf{C1}, (b)~\textbf{C2}, (c)~\textbf{C3},
    (d)~\textbf{C4}}{Averaged mean $C_p$ as a function of angle from
    the front stagnation point, (a)~\textbf{C1}, (b)~\textbf{C2}, (c)~\textbf{C3},
    (d)~\textbf{C4}}
  \label{fig:Cpmean}

\end{sidewaysfigure}

Figure~\ref{fig:Cp0} presents time-averaged surface pressure
distributions against $\theta$ from the front stagnation
point for the middle circular cylinder, taken from
the first column to the fourth column, respectively. The results
of \citeauthor{Shim1985}\cite{Shim1985} are shown for comparison. 

The surfaces pressures are presented in terms of the coefficient of
pressure
\beq
\overline{C}_p = \frac{\left< p \right>_{T} -
  p_{ref}}{q_{ref}}
\label{eq:cp}
\eeq
where $\left< p \right>_{T}$ denotes an ensemble average across
the spanwise direction for all the sampling points on the
cylinder surface over the
sampling time interval $T$, though the vortex shedding does
not necessarily occur in phase over the whole spanwise direction. The time-averaged boundary
layers on either side of each circular cylinder are assumed
to be symmetrical. $q_{ref}$ is the dynamic
pressure in terms of free stream velocity $u_o$ and fluid
density $\rho$, which is given by
\beq 
 q_{ref} = \frac{1}{2}\rho u_g^2 
\eeq
To make $\overline{C}_p$ equal to unit at the front
stagnation point for every cylinder, the corresponding
static pressure $p_{ref}$ is calculated according to
equation \ref{eq:cp} first, $\overline{C}_p$ is hence
determined around the cylinder surface. This procedure was
also used in the work of \citeauthor{Shim1985}\cite{Shim1985} for calculating
$\overline{C}_p$. Hence, in view of the transverse pitch ratio
$S_T = 2.67$ and in terms of the continuity equation
$u_g/u_o = S_T/(S_T - d)$, one obtains the corresponding
converting factor for the related quantities.

Very good agreements for the time-averaged surface pressure
distribution around the four cylinders are observed between
the LES calculations and the experimental
measurements of \citeauthor{Shim1985}\cite{Shim1985} among the four figures of
Figure \ref{fig:Cpmean}. Other quantities, for example, the
r.m.s pressure distribution and vortex shedding frequency
are also very comparable. They shall be shown in later
figures in this paper. For the cylinder from the first column,
note that the LES data in figure \ref{fig:Cp0} contain a
kink near $\theta = 85^{o}$, which indicates the presence of
the laminar boundary layer separating from the upper and
lower surface of the cylinder. This transition region from
the experimental data of \citeauthor{Shim1985}\cite{Shim1985} is not as readily
perceived as in the LES computation, in that the
measurements were taken in 10-degree increments from the
forward stagnation point to the opposite side of the
cylinder.  For the discernible wiggle from the present
calculations in the range of $\theta \in \left[80, 120
\right] $, the likely reason is entrainment of shear layer
fluid on to the cylinder surface owing to the interference
from the close arrangement of cylinders.

To the best of the authors' knowledge, there is so far no
information available on the pressure distribution around
the surfaces of cylinders in a tube bank from LES. It is of
interest, thus, to show mean pressure distribution around
the surface of the downstream cylinders in terms of the
equation \ref{eq:cp}, and to further compare the results
measured by \citeauthor{Shim1985} \cite{Shim1985}. As far as the positive values
of base $\overline{C}_p$ obtained from the downstream
cylinders are concerned, it also results from the definition
of $\overline{C}_p$ in this work. It can be observed
that the results from the two distinct approaches are very
comparable across the figure \ref{fig:Cp1},
\subref*{fig:Cp2}, \subref*{fig:Cp3}. Because of the wake
from the first column of cylinders which impinges upon the
downstream second column of cylinders, a rise of mean pressure
value is to be expected within the windward side. In
particular, as can be observed from the figure \ref{fig:Cp0} and
\ref{fig:Cp1}, they display distinct shapes for mean
pressure distribution. The rise of mean pressure
distribution is clearly discernible within the range of
$\theta \in \left[0^o, 40^o\right]$ in
figure~\ref{fig:Cp1}. It is interesting to note that the two
peaks lie nearly to the same position around $\theta = 40^o$
in Figure~\ref{fig:Cp1}. Moreover, it is worthwhile noting
that the difference from the pressure of front stagnation
point and the base pressure is reduced significantly
compared to the corresponding cylinder from the first
column. This is attributed to the turbulence level of
approaching flow since it is located within the wake. In
contrast to figure~\ref{fig:Cp1}, the rise is not observed
for$\overline{C}_P$ from the third and forth column in
figure~\ref{fig:Cp2},\subref*{fig:Cp3}. This can be
explained as the wake from downstream cylinders is much
narrow and more mixed than the one behind the first column.

\begin{sidewaysfigure}
\vspace{6.0in}   
\centering
  \subfloat{\label{fig:Cp0_rms}\includegraphics[trim=0.00in
    0.00in 0.00in 0.00in,clip=true,scale=1.0, width = 0.4\textwidth]{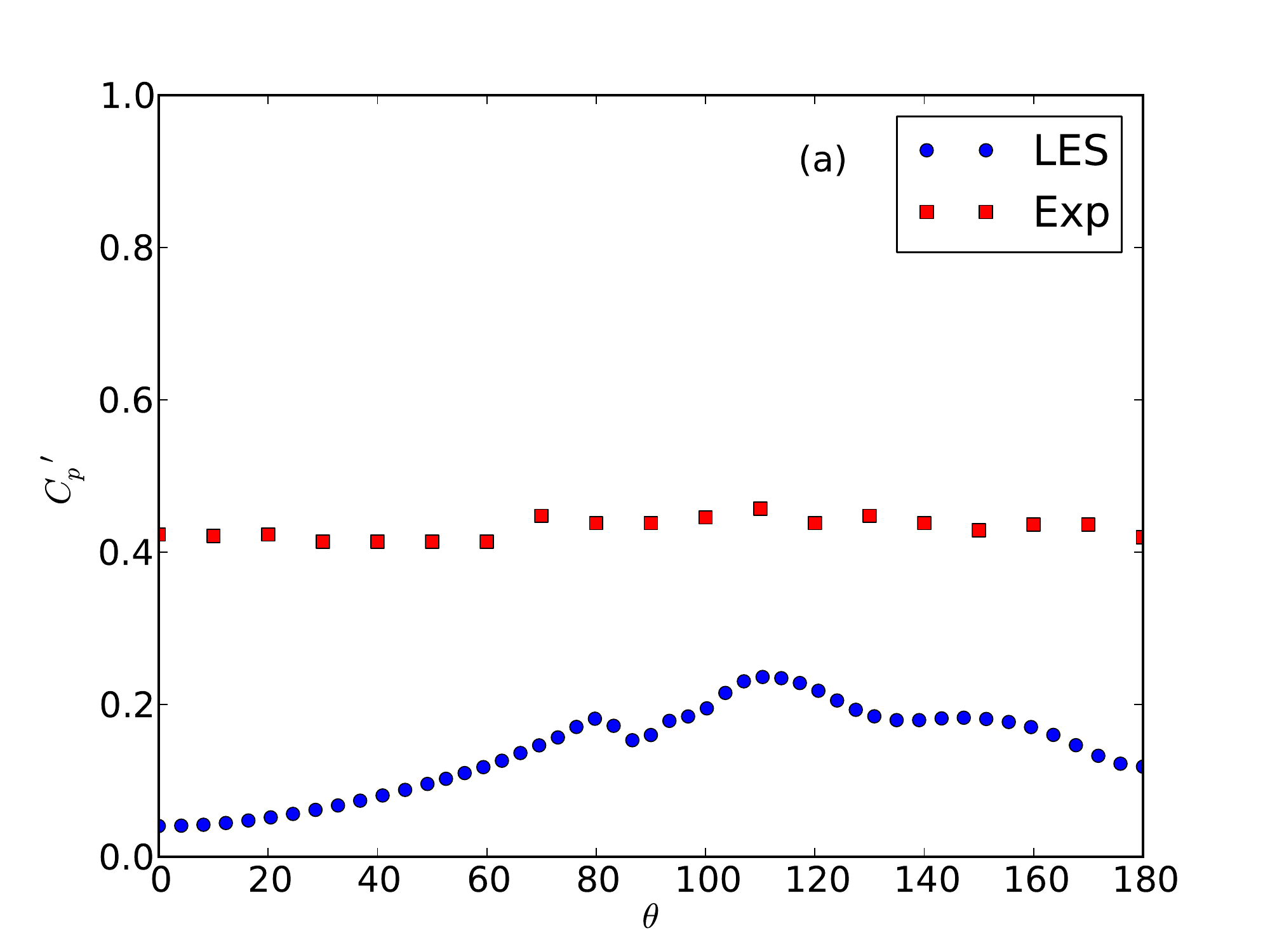}} 
  \subfloat{\label{fig:Cp1_rms}\includegraphics[trim=0.00in
    0.00in 0.00in 0.00in,clip=true,scale=1.0, width =
    0.4\textwidth]{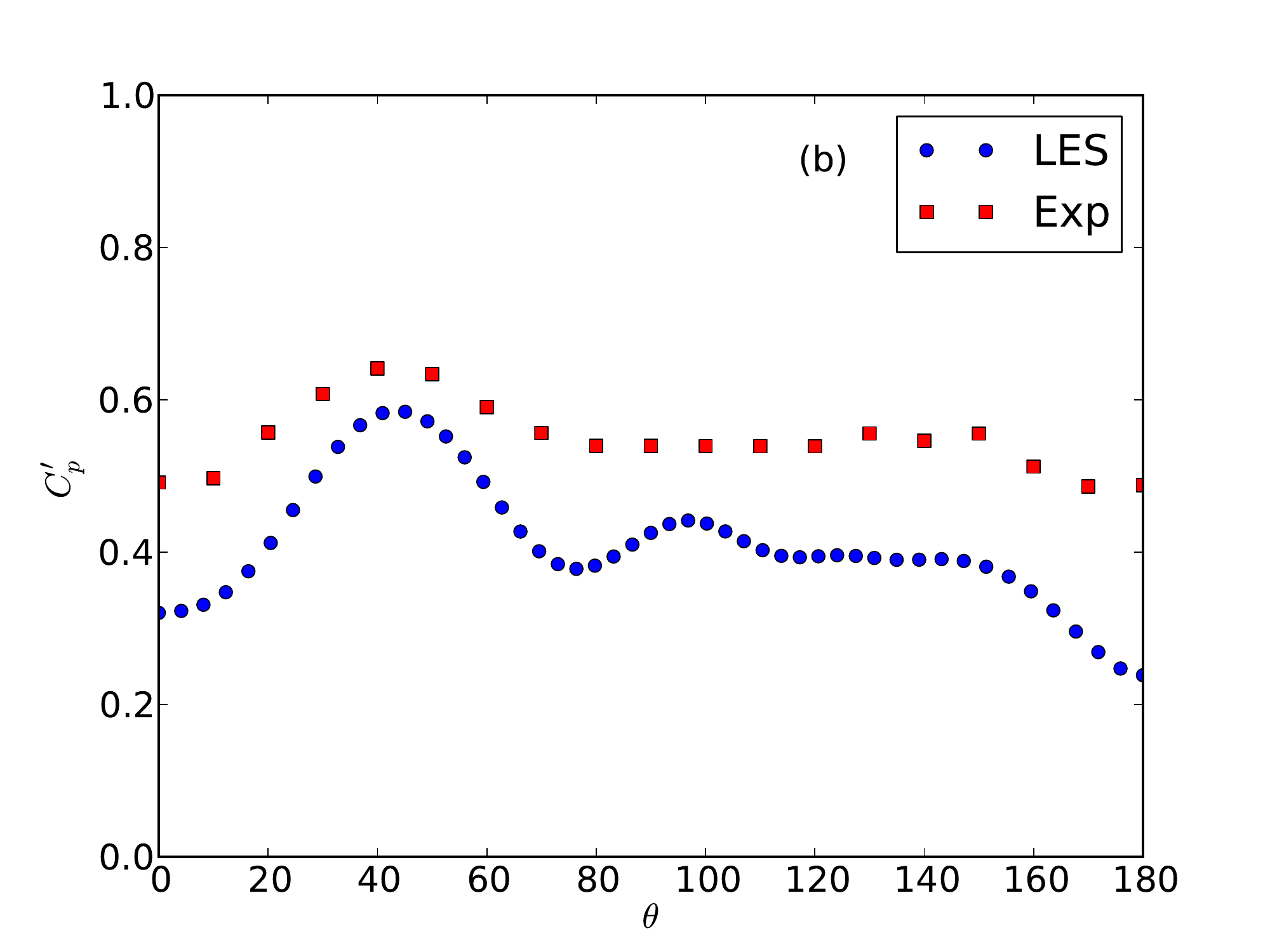}} \\
  \subfloat{\label{fig:Cp2_rms}\includegraphics[trim=0.00in
    0.00in 0.00in 0.00in,clip=true,scale=1.0, width = 0.4\textwidth]{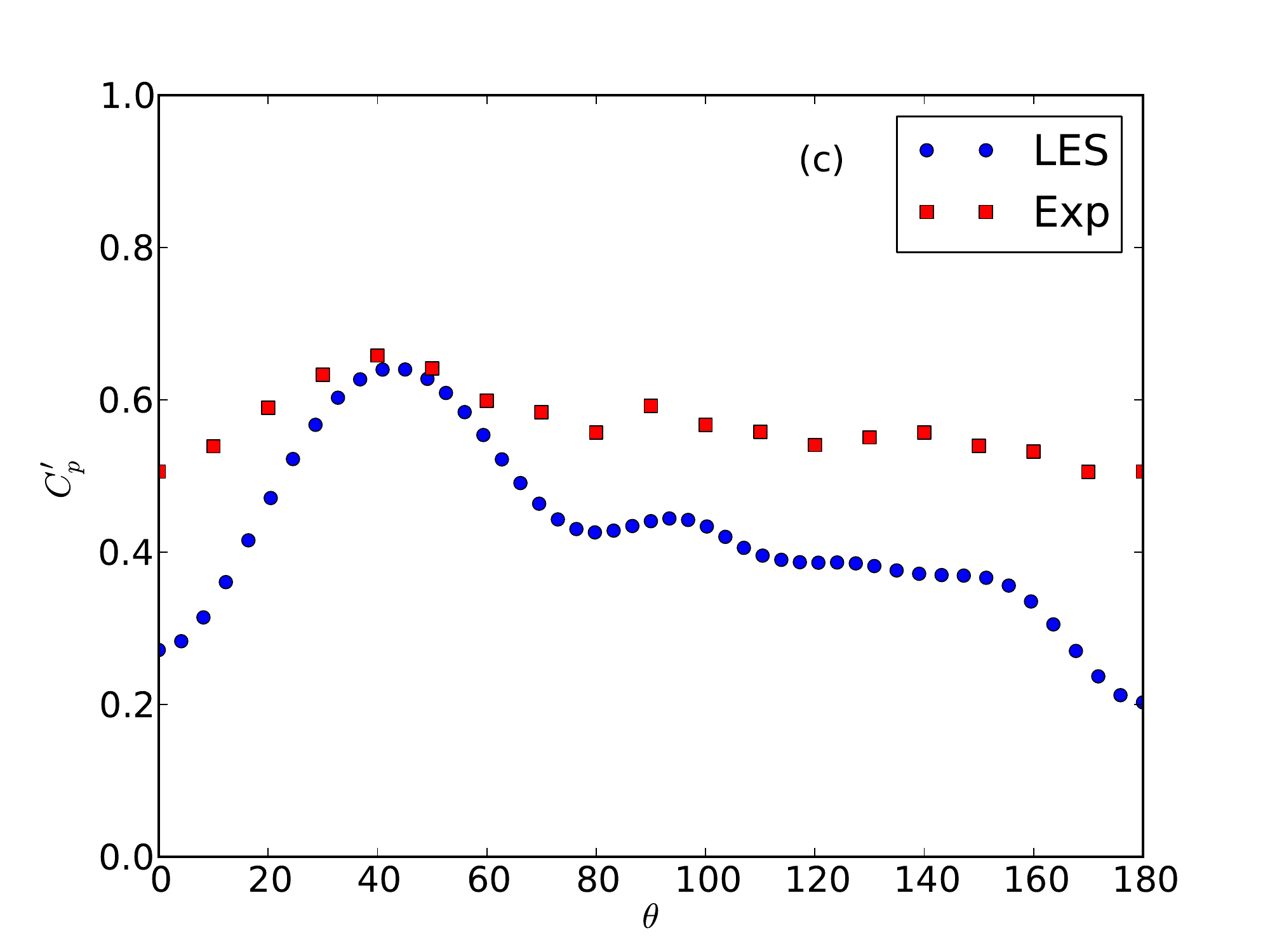}}  
  \subfloat{\label{fig:Cp3_rms}\includegraphics[trim=0.00in
    0.00in 0.00in 0.00in,clip=true,scale=1.0, width =
    0.4\textwidth]{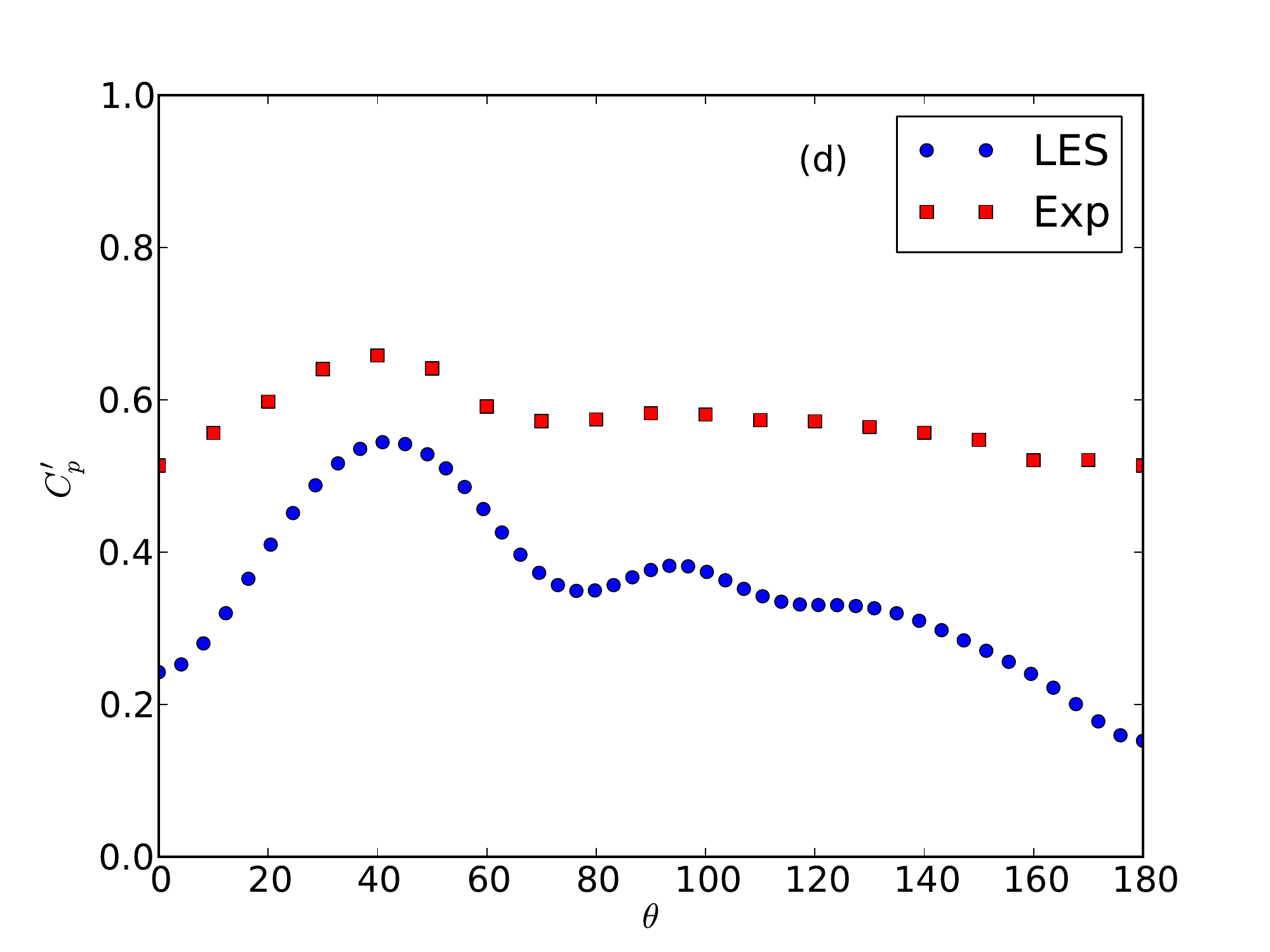}} 
  
  \caption[surface fluctuating pressure distribution as a function of angle from
    the front stagnation point over 30 non-dimensional time
    units,(a)~\textbf{C1}, (b)~\textbf{C2}, (c)~\textbf{C3},
    (d)~\textbf{C4}]{surface fluctuating pressure distribution as a function of angle from
    the front stagnation point over 30 non-dimensional time
    units,(a)~\textbf{C1}, (b)~\textbf{C2}, (c)~\textbf{C3},
    (d)~\textbf{C4}} 
  \label{fig:Cprms}

\end{sidewaysfigure}

The r.m.s value of pressure distribution around the surfaces
of the four cylinders are shown in the four
figures~\ref{fig:Cp0_rms},\subref*{fig:Cp1_rms},\subref*{fig:Cp2_rms},\subref*{fig:Cp3_rms}
along with the \citeauthor{Shim1985}\cite{Shim1985}'s data. First to
note is that the
pressure fluctuates more than $50\%$ for the downstream
cylinders. This indicates that instantaneous surface
pressure different from the time-averaged value
significantly and further demonstrates that the URANS
methodology is not suitable for the present work.
Figure~\ref{fig:Cp0_rms} exhibits relatively high and
uniform values of fluctuating pressure distributions around
the first cylinder from the findings of \citeauthor{Shim1985} \cite{Shim1985}. One
must suspect this according to the work of
\citeauthor{norberg2003fluctuating}\cite{norberg2003fluctuating} at a comparable Reynolds
number that $C_P^{\prime}$ exhibits a very low level at the
frontal stagnation line ($\theta = 0^o$). Second, it can be
observed that the general trend is in reasonably good
agreement with the measurements of \cite{Shim1985} except for
the first column cylinder. Moreover, the position of first peak
as shown in the figure~\ref{fig:Cp0_rms} corresponds to the
same angle in the figure~\ref{fig:Cp0} that indicates the
tripping of laminar boundary layer separation. The second
peak after the shoulder of the cylinder results from the
reattachment of boundary layer separation on the surface.

Table~\ref{tab:Cprms} shows a comparison of r.m.s value and
maximum value of pressure fluctuations from the present LES
computation with the experimental values of
\citeauthor{Shim1985}\cite{Shim1985} and
\citeauthor{norberg2003fluctuating}\citeauthor{norberg2003fluctuating}. It
can be observed that the values from the present
calculations match very well with the measurements,
especially for the angular position within the windward side
at which the maximum r.m.s value of fluctuating pressure
occurs. In addition, one interesting point is that the width
of wakes from the second, third and fourth column cylinder
is very close.  The maximums on downstream cylinders are
caused by the impingement of shedding-vortex from the
upstream cylinders.  The low r.m.s values of pressure
fluctuation compared with experimentally measured ones on
the leeward side result from the relatively weak wake
predicted by the present LES with wall-layer modelling.  It
is also worthwhile emphasizing that the calculated results
at $\theta = 90^o$ are significantly higher that the value
at a comparable $Re = 10k$ compiled in \citeauthor{norberg2003fluctuating}
\cite{norberg2003fluctuating} for a single circular
cylinder.

\begin{table}\small
\centering
\begin{tabular*}{0.95\textwidth}{p{5cm}p{2cm}ll}
  \toprule                                                
  
  Case                                       & \qquad & Maximum of $C_p^{\prime}$ & $C_p^{\prime}(90^o)$ \\
  \midrule
  \multirow{4}{*}{Present LES~($Re_g = 15270$)}        & C1
  & 0.236($110^o$)            & 0.159   \\
                                             & C2     & 0.584($40.4^o$)           & 0.425                \\
                                             & C3     & 0.640($40.4^o$)           & 0.441                \\
                                             & C4     & 0.544($36.7^o$)           & 0.377                \\
  \midrule
  \multirow{4}{*}{Experiments~\citep{Shim1985}} 
                                             & C1     & 0.457($110^o$)            & 0.438                \\
                                             & C2     & 0.641($40^o$)             & 0.539                \\
                                             & C3     & 0.658($40^o$)             & 0.592                \\
                                             & C4     & 0.658($40^o$)             & 0.582                \\
  \midrule
  Experiments~\citep{norberg2003fluctuating} Single cylinder
  $Re = 10k$ & 
                   & 0.292  & 0.282                                            \\
\bottomrule
\end{tabular*} 
\caption{Comparison of results for r.m.s pressure
  distribution $C_p^{\prime}$}
\label{tab:Cprms}
\end{table}

Finally, judging from the shape of
mean pressure distribution around the surface in the two
figures~\ref{fig:Cpmean},~\ref{fig:Cprms},
it stands to reason that the present calculation is capable
of accurately predicting the pattern or dynamics of flow across tube-banks.

\subsection{Drag and lift coefficients}
\begin{figure}[h!tbp]
  \centering
  \subfloat{\label{fig:Cp0_rms}\includegraphics[trim=0.00in
    0.00in 0.00in 0.00in,clip=true,scale=1.0, width =
    0.9\textwidth]{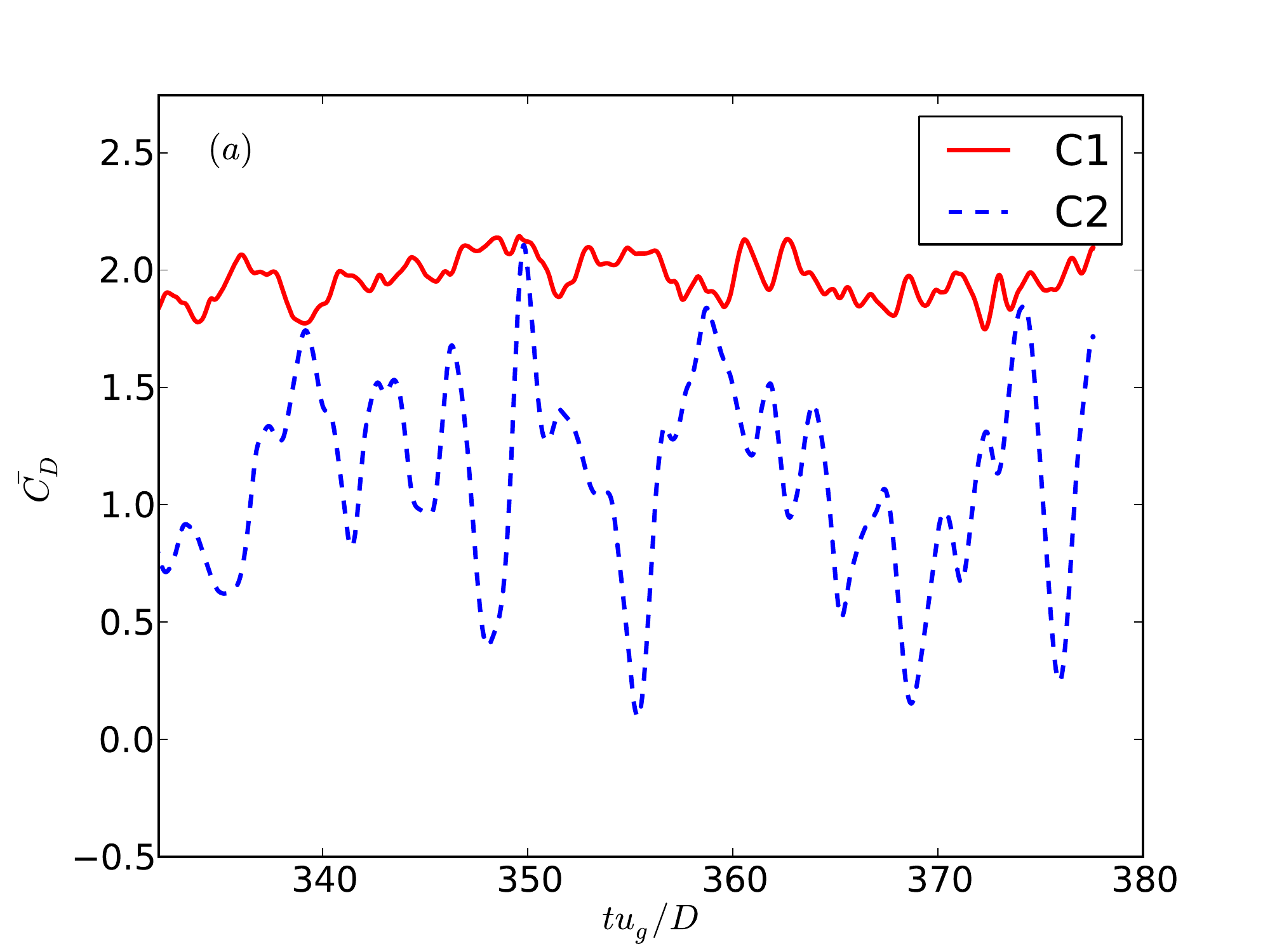}}\\
  \subfloat{\label{fig:Cp0_rms}\includegraphics[trim=0.00in
    0.00in 0.00in 0.00in,clip=true,scale=1.0, width =
    0.9\textwidth]{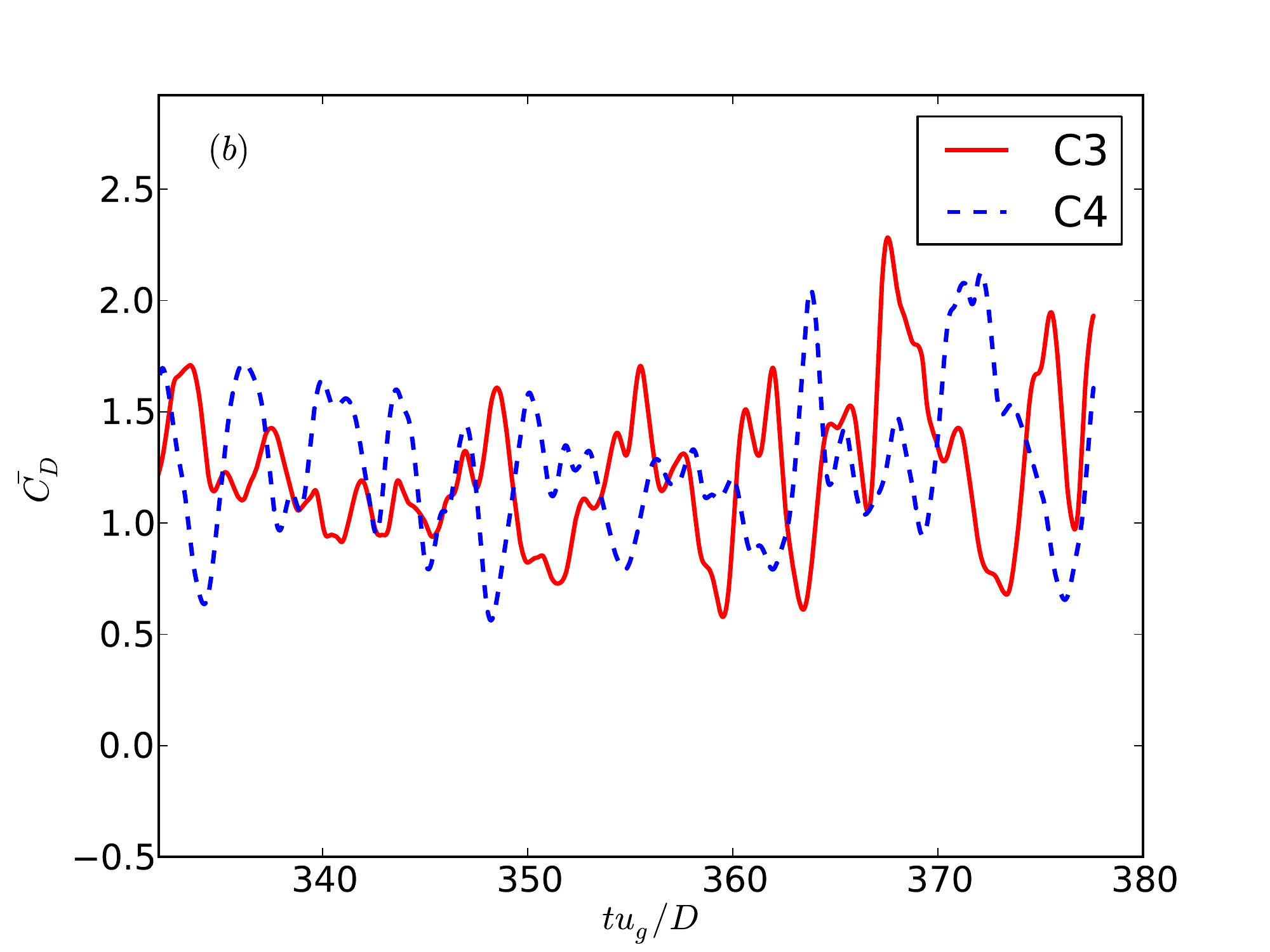}}
   \caption[Time history of $\overline{C}_D$,
     (a)~\textbf{C1,C2} (b)~\textbf{C3,C4}]{Time history of $\overline{C}_D$,
     (a)~\textbf{C1,C2} (b)~\textbf{C3,C4}}
  \label{fig:Cd1234}
\end{figure}

To further validate the present study with experiments,
table~\ref{tab:CdCl} summarizes the flow parameters
concerning $\overline{C}_D$ and $C_L^{\prime}$ along with 
experimental measurements. The coefficient of
mean drag per unit span is defined by:
\beq
\overline{C}_D = \frac{\overline{F}_D}{ld\rho u_g^2/2}
\label{eq:meancd}
\eeq 
where $l$ denotes the spanwise length of the cylinder;
$\overline{F}_D$ denotes the form drag force caused by the
surface pressure distribution through ignoring the viscous
drag force, which is obtained by an integration of
mean pressure distribution around the cylinder. Thus,
$\overline{C}_D$ is given by  
\beq
\overline{C}_D =
\int_{0^o}^{180^o}\overline{C}_P cos(\theta)\mathrm{d}\theta.
\eeq
It is evident that the results of $\overline{C}_D$ predicted
by the present LES study agree favorably well with the
experimental measurements (\citep{Shim1985}) except
under-predict $\overline{C}_D$ for the second column
cylinder; the magnitudes of $C_L^{\prime}$ obtained from
this work shows reasonable agreement with experimentally
measured values except for the first column cylinder due to
the reason discussed before. The results for
$\overline{C}_D$ and $C_L^{\prime}$ are also interpreted in
terms of with the free stream velocity $u_o$, which is based
on the conversion factor discussed in section~\ref{sec:spd}
and clearly a significant variable concerning
$\overline{C}_D$ and $C_L^{\prime}$ as shown in
table~\ref{tab:CdCl}. Through interpreting this way, the
drag experienced by the first column cylinder is increased
considerably. Similar observations apply to the rest of
downstream cylinders. But the increment for the first
column cylinder is distinct from the remaining
ones. Comparing the value $\overline{C}_D = 1.941$ in terms
of $Re_g = 15270$ with $\overline{C}_D = 1.185$
(\citep{schlichting2000boundary}) for a comparable Reynolds
number, it can be observed that $\overline{C}_D$ for the
first column cylinder predicted in this LES study is
considerably higher that the value for a unconfined single
smooth circular cylinder. This can be explained that the
distinct discrepancy interpreted through $u_o$ is
undoubtedly a consequence of a higher pressure coefficient
which results from higher separation velocities
in confined flow situations (\citep{richter1976fluctuating})
as shown in figure~\ref{fig:Cd1234}.

In the light of the foregoing discussion it becomes evident
that the $\overline{C}_D$ for the downstream cylinders would
be much higher than an unconfined circular cylinder as
well. Nevertheless, compared with the standard value
$\overline{C}_D = 1.185$ (\citep{schlichting2000boundary}),
table~\ref{tab:CdCl} shows comparable values for
$\overline{C}_D$. Thus, it seems reasonable that the transverse pitch
ratio $S_T = P_T/D = 2.67$ does not give rise to the same
effect on the drag for the downstream cylinder as for the
first column. However, in the present study, the approaching
stream for different column cylinders is of different
turbulence level which brings about appreciable effects on
the vortex shedding and drag force. For example, the free
stream for the first column cylinder is assumed
uniform. Whilst the downstream cylinders encounter significantly high
turbulence level because they lie in the wake. Hence, the
blockage ration and the turbulence level of approaching
stream, two conflicting factors, result in a drag
coefficient $\overline{C}_D$ for the downstream cylinder
which is not significant different from an unconfined
circular smooth cylinder. This is demonstrated in
figure~\ref{fig:Cd1234} which shows $\overline{C}_D$ for
cylinder C2, C3, C4 is significant lower than the value of
C1.

\begin{table}[t!]\small
\centering
\begin{tabularx}{0.95\textwidth}{p{3.6cm}Xp{1.9cm}p{1.7cm}p{1.7cm}p{1.9cm}}
  \toprule                                               
  \multirow{2}{*}{Case} & & $\overline{C}_D$ & $C_L^{\prime}$ &
  $\overline{C}_D$ & $C_L^{\prime}$ \\ \cline{3-6}
& & \multicolumn{2}{l}{Based on $u_g$} &
  \multicolumn{2}{l}{Based on $u_o$} \\
  \midrule
  \multirow{4}{*}{Present LES} 
& C1 & 0.767 & 0.228 & 1.941 & 0.579  \\
& C2 & 0.404 & 0.655 & 1.022 & 1.656  \\
& C3 & 0.454 & 0.650 & 1.146 & 1.645  \\
& C4 & 0.464 & 0.507 & 1.174 & 1.284  \\ 
   \midrule
  \multirow{3}{*}{Experiments}
&    &  \multicolumn{2}{l}{Estimated} &  \multicolumn{2}{l}{Estimated} \\ \cline{3-6} 
& C1 & 0.799 & 0.05-0.08 & 2.022 & 0.127-0.202  \\
\citep{Shim1985}& C2 & 0.324 & 0.55-0.65 & 0.820 & 1.391-1.645  \\
& C3 & 0.465 & 0.60-0.70 & 1.176 & 1.518-1.771  \\
& C4 & 0.476 & 0.52-0.60 & 1.204 & 1.316-1.518  \\
   \midrule
   \multirow{1}{*}{Emp.correlation}
&    &  \multicolumn{2}{l}{$Re = 15270$} &  \multicolumn{2}{l}{$Re=9600$} \\ \cline{3-6}
 \citep{norberg2003fluctuating}  & & & 0.520 & & 0.520 \\
   \midrule
   \multirow{1}{*}{Experiments~}
   & & & &\multicolumn{2}{l}{Blockage ratio = 1/4}   \\ \cline{3-6}
 \citep{richter1976fluctuating}  & & & &1.35-1.40 &0.80-0.90  \\
  \bottomrule
\end{tabularx} 
\caption{Comparison of results for $\overline{C}_D$ and $C_L^{\prime}$}
\label{tab:CdCl}
\end{table}

\begin{figure}[h!tbp]
  \centering
  \subfloat{\label{fig:Cp0_rms}\includegraphics[trim=0.00in
    0.00in 0.00in 0.00in,clip=true,scale=1.0, width =
    0.9\textwidth]{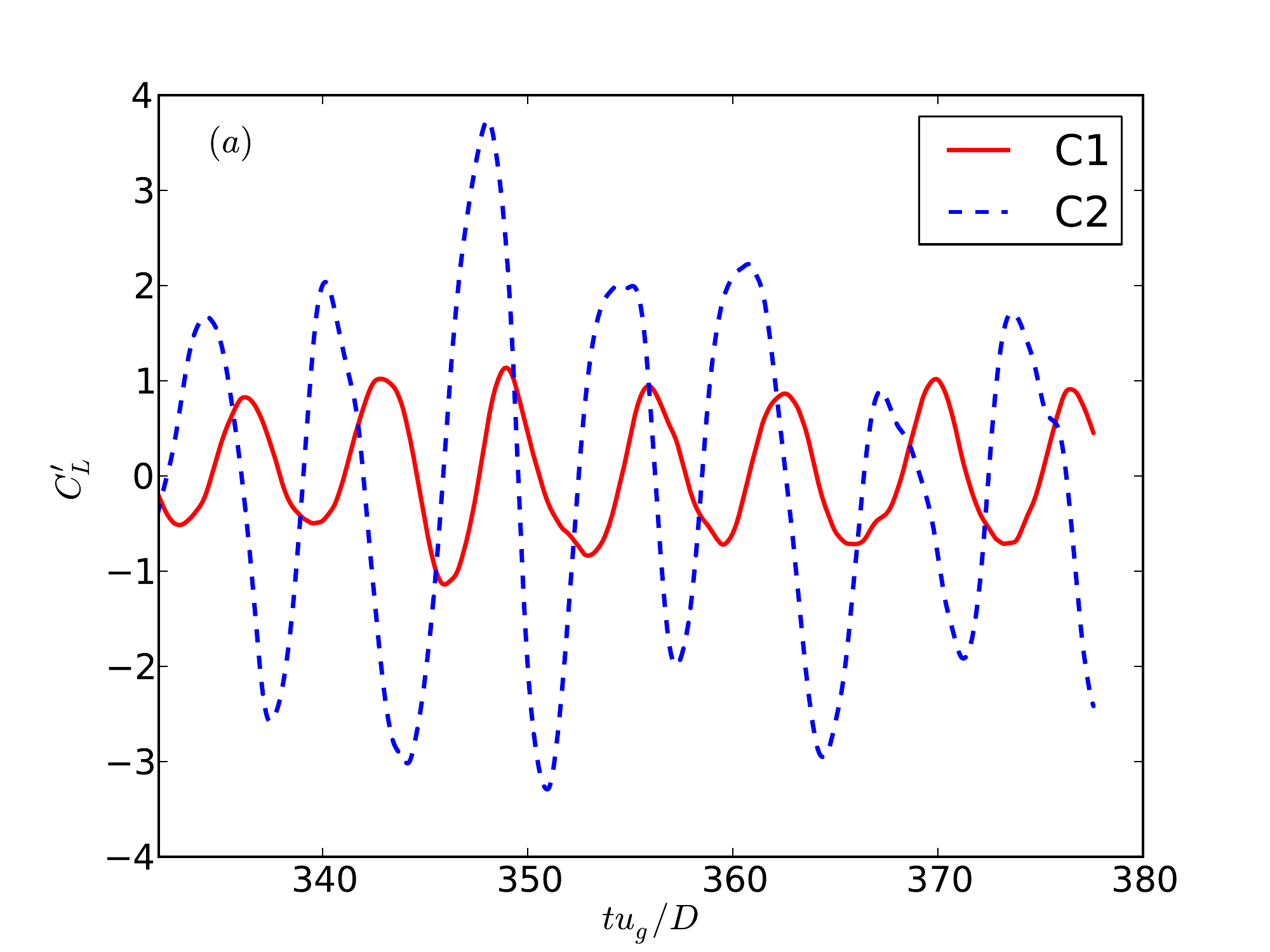}}\\
  \subfloat{\label{fig:Cp0_rms}\includegraphics[trim=0.00in
    0.00in 0.00in 0.00in,clip=true,scale=1.0, width =
    0.9\textwidth]{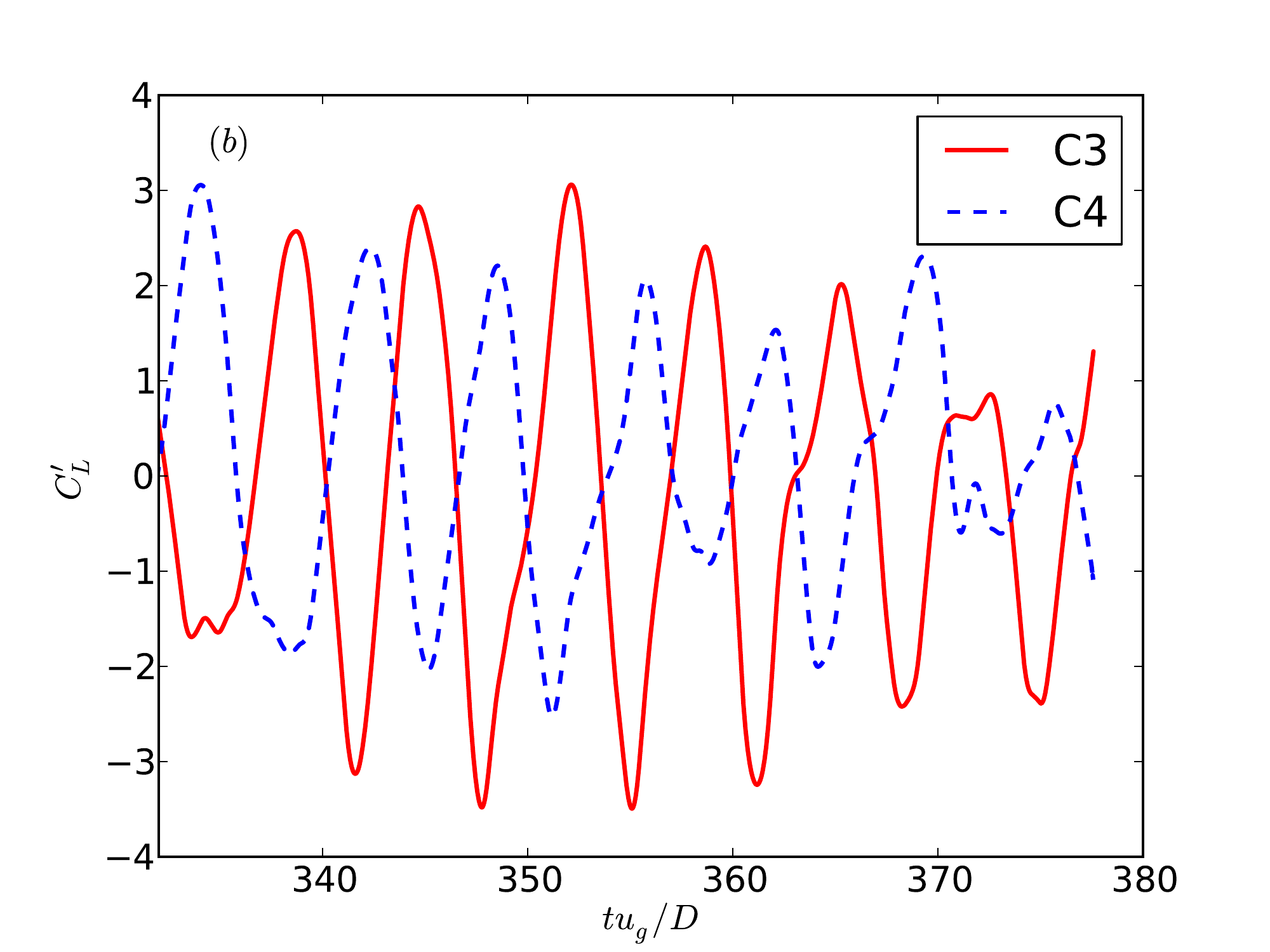}}
   \caption[Time history of $C_L^{\prime}$ , (a)~\textbf{C1,C2} (b)~\textbf{C3,C4}]{Time history of $C_L^{\prime}$ , (a)~\textbf{C1,C2} (b)~\textbf{C3,C4}}
  \label{fig:Cl1234}
\end{figure}

In selecting extra experimental data, the focus is given to
those measured on a single circular smooth cylinder in free
stream flow or in confined flows with a comparable blockage
ratio to the present configuration. 
The data for $C_L^{\prime}$ from \citeauthor{norberg2003fluctuating}\cite{norberg2003fluctuating} are
determined by the following correlation
\beq
C_L^{\prime} = 0.52 - 0.06 \times [log(Re/1600)]^{-2.6}
\qquad (5.4\times10^3 < Re < 2.2\times10^5)
\eeq
which covers the upper bound of sub-critical Reynolds number
range. $C_L^{\prime}$ does not display much variation when
the Reynolds number remains below the critical
value. \citeauthor{richter1976fluctuating}\cite{richter1976fluctuating}'s data , which are
extrapolated from their experimental observations performed
at a smooth circular cylinder in a wind-tunnel with a
blockage ratio of $1/4$, are included for further
comparisons. It can observed that, when interpreting
$C_L^{\prime}$ in terms of $u_o$, the values of
$C_L^{\prime}$ for downstream cylinders are significantly
higher than that of an unconfined circular cylinder and do
not fall in the scatter region of an confined circular
cylinder (\cite{richter1976fluctuating}). This indicates that
the vortex shedding from downstream cylinders may be
augmented further by the feature of approaching wake
turbulence from upstream cylinders. Consequently, it yields
considerably higher values of $C_L^{\prime}$ on downstream
cylinders as presented in figure~\ref{fig:Cl1234} in which
$C_L^{\prime}$ is interpreted with the free stream
velocity $u_o$.

Moreover, two interesting points can be derived from
figure~\ref{fig:Cd1234} and \ref{fig:Cl1234}. First, as far
as the fluctuating magnitudes of the two variables are
concerned, it can be observed that the first column displays
smaller values than downstream cylinders. The second
observation is that the maximum fluctuation of
$C_L^{\prime}$ is achieved on the second or the third column
cylinder. A similar phenomena was reported by \citeauthor{liang2007large}.
\cite{liang2007large}.

\subsection{Shear-layer instability and vortex shedding}

\begin{sidewaysfigure}
\vspace{6.0in}
\centering
  \subfloat{\label{fig:C1vector}\includegraphics[trim=0.8in
    0.8in 0.8in 0.8in,clip=true,scale=1.0, width =
    0.4\textwidth]{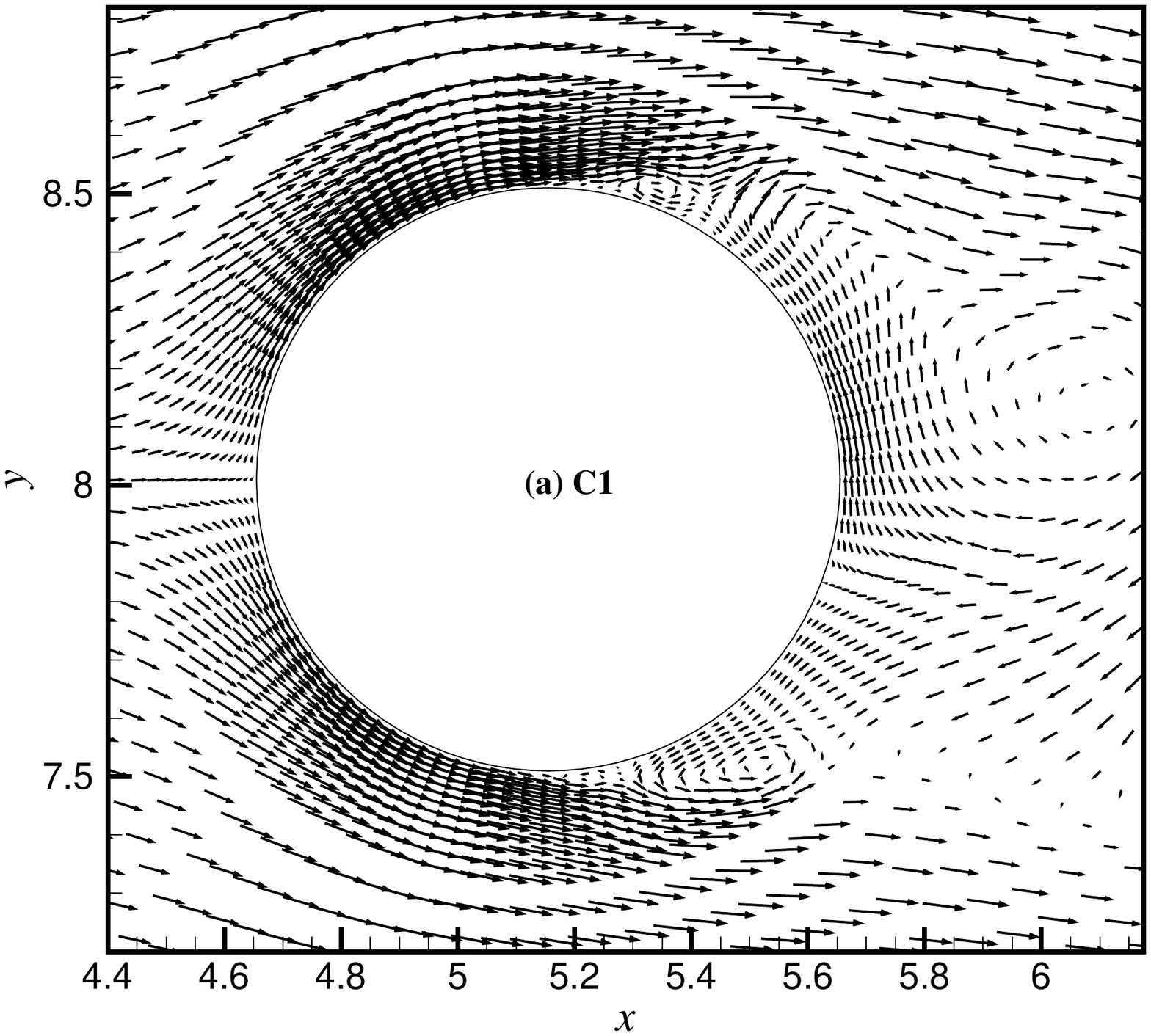}}
  \subfloat{\label{fig:C2vector}\includegraphics[trim=0.8in
    0.8in 0.8in 0.8in,clip=true,scale=1.0, width =
    0.4\textwidth]{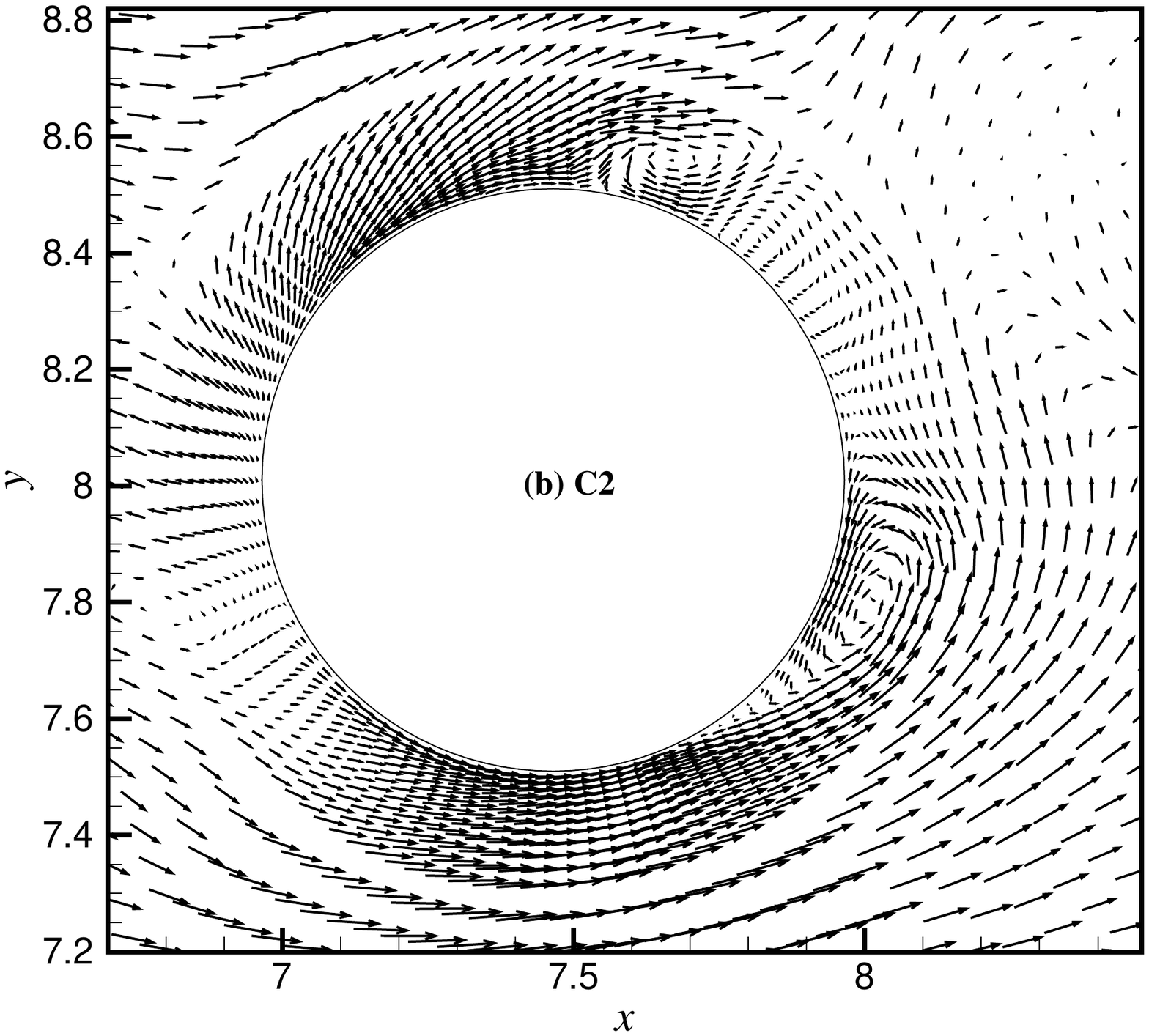}} \\
  \subfloat{\label{fig:C3vector}\includegraphics[trim=0.8in
    0.8in 0.8in 0.8in,clip=true,scale=1.0, width =
    0.4\textwidth]{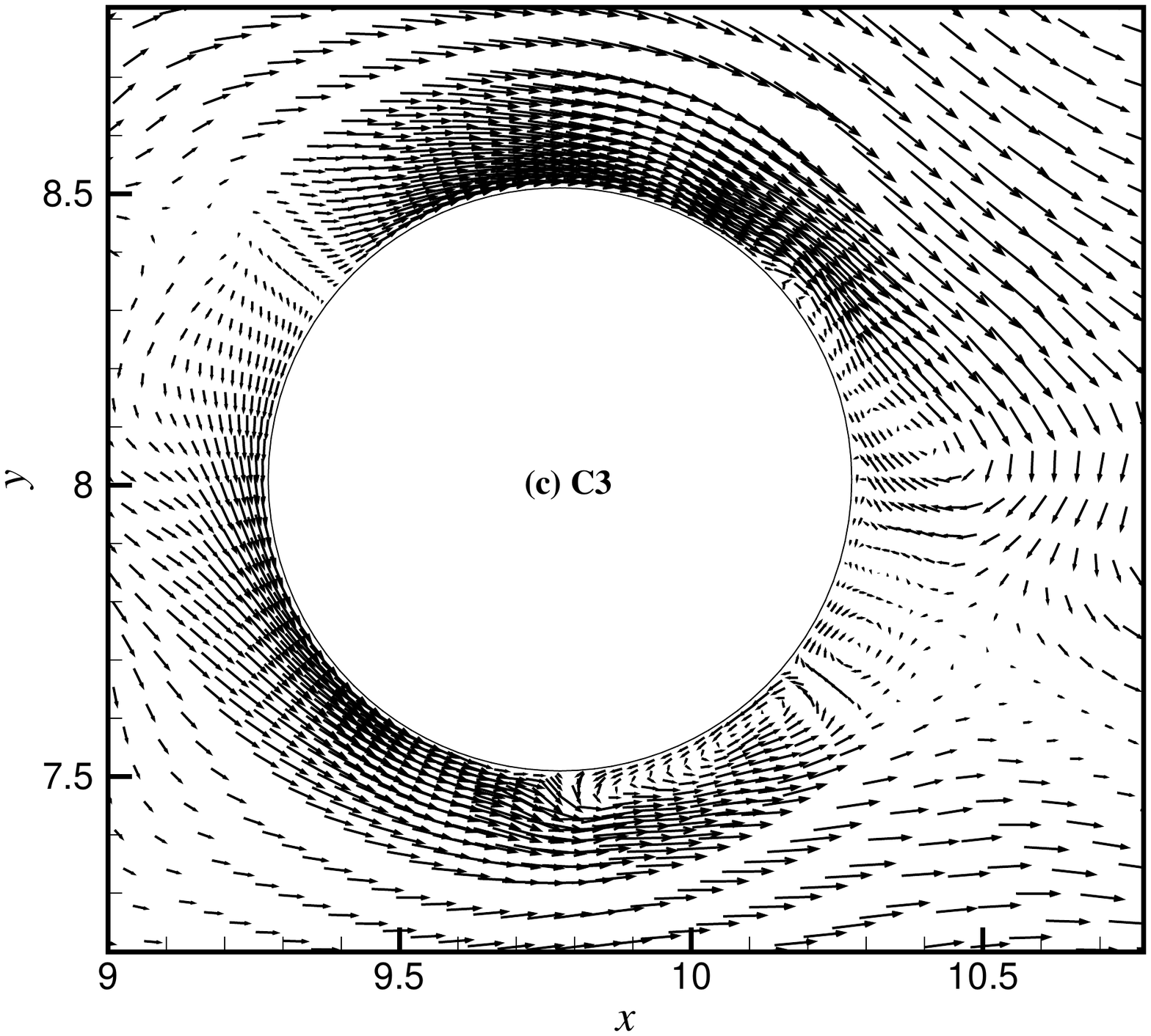}}
  \subfloat{\label{fig:C4vector}\includegraphics[trim=0.8in
    0.8in 0.8in 0.8in,clip=true,scale=1.0, width =
    0.4\textwidth]{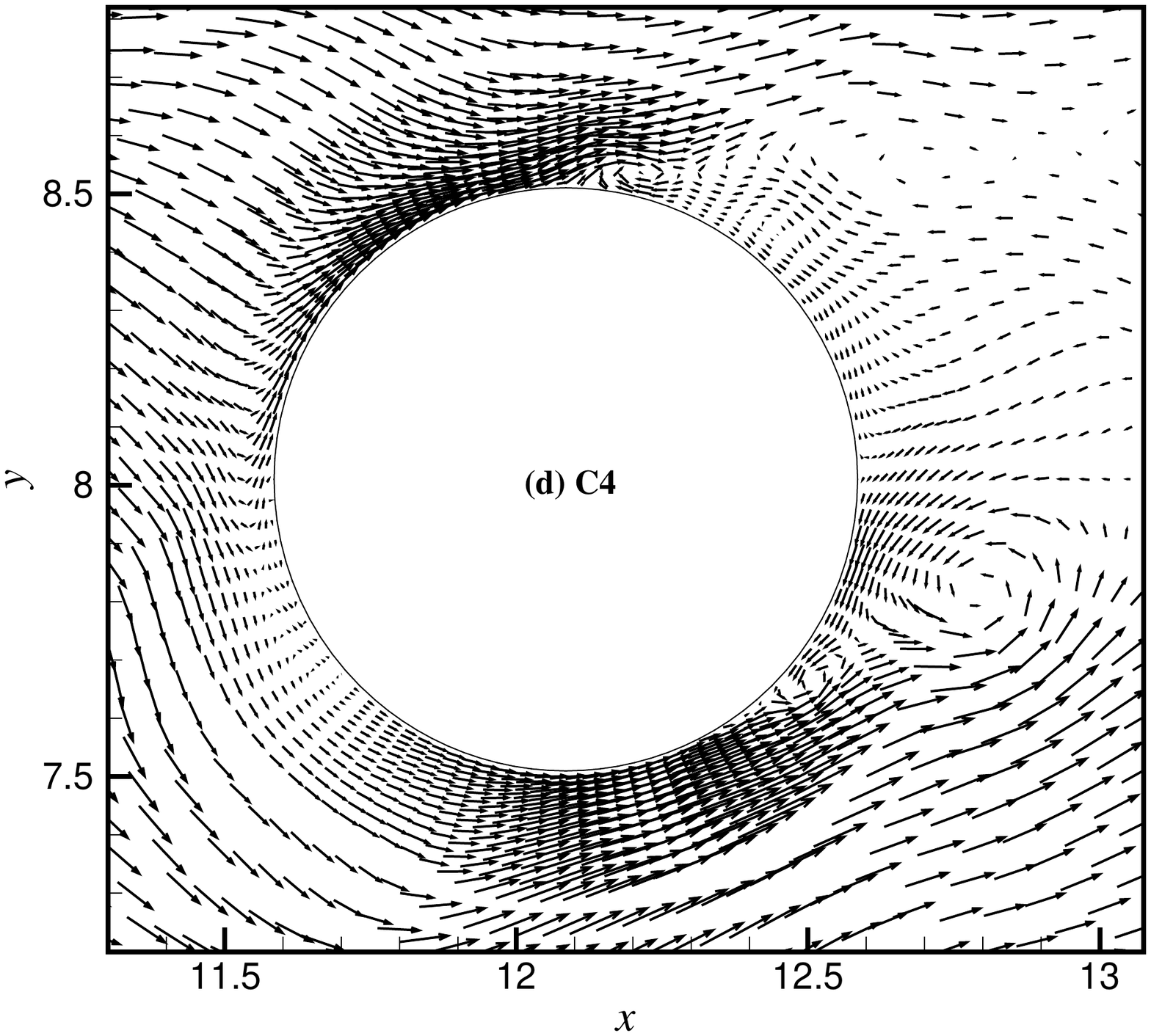}}
    \caption[Instantaneous velocity vector map in the middle
    plane of flow domain (a)~\textbf{C1}, (b)~\textbf{C2}, (c)~\textbf{C3},
    (d)~\textbf{C4}]{Instantaneous velocity vector map in the middle
    plane of flow domain (a)~\textbf{C1}, (b)~\textbf{C2}, (c)~\textbf{C3},
    (d)~\textbf{C4}}
  \label{fig:C1234vector}

\end{sidewaysfigure}

Figure~\ref{fig:C1234vector} presents close-up views of an
instantaneous velocity vector map in the middle plane of the
flow domain around the four cylinders C1, C2, C3, C4. In
accordance with the results of previous researchers
(\cite{bloor1964transition}), in the sub-critical
regimes, the separating shear layers becomes turbulent. In
figure~\ref{fig:C1vector}, it can be observed that
small-scale vortexes are being formed in the shear layers
behind C1. Nevertheless, such small vortexes appear not to be
formed behind the downstream cylinders as shown in
figure~\ref{fig:C2vector} \subref*{fig:C3vector}
\subref*{fig:C4vector}. This can be explained that the
approaching stream for the downstream cylinders, the wake of
upstream cylinders, is of particularly high turbulence level,
so that there is no transition that is closely connected
with the vortex formation in the shear layer.

\begin{figure}[h!tbp]
  \centering
  \subfloat{\label{fig:Cd0_psd}\includegraphics[trim=0.00in
    0.00in 0.00in 0.00in,clip=true,scale=1.0, width =
    0.9\textwidth]{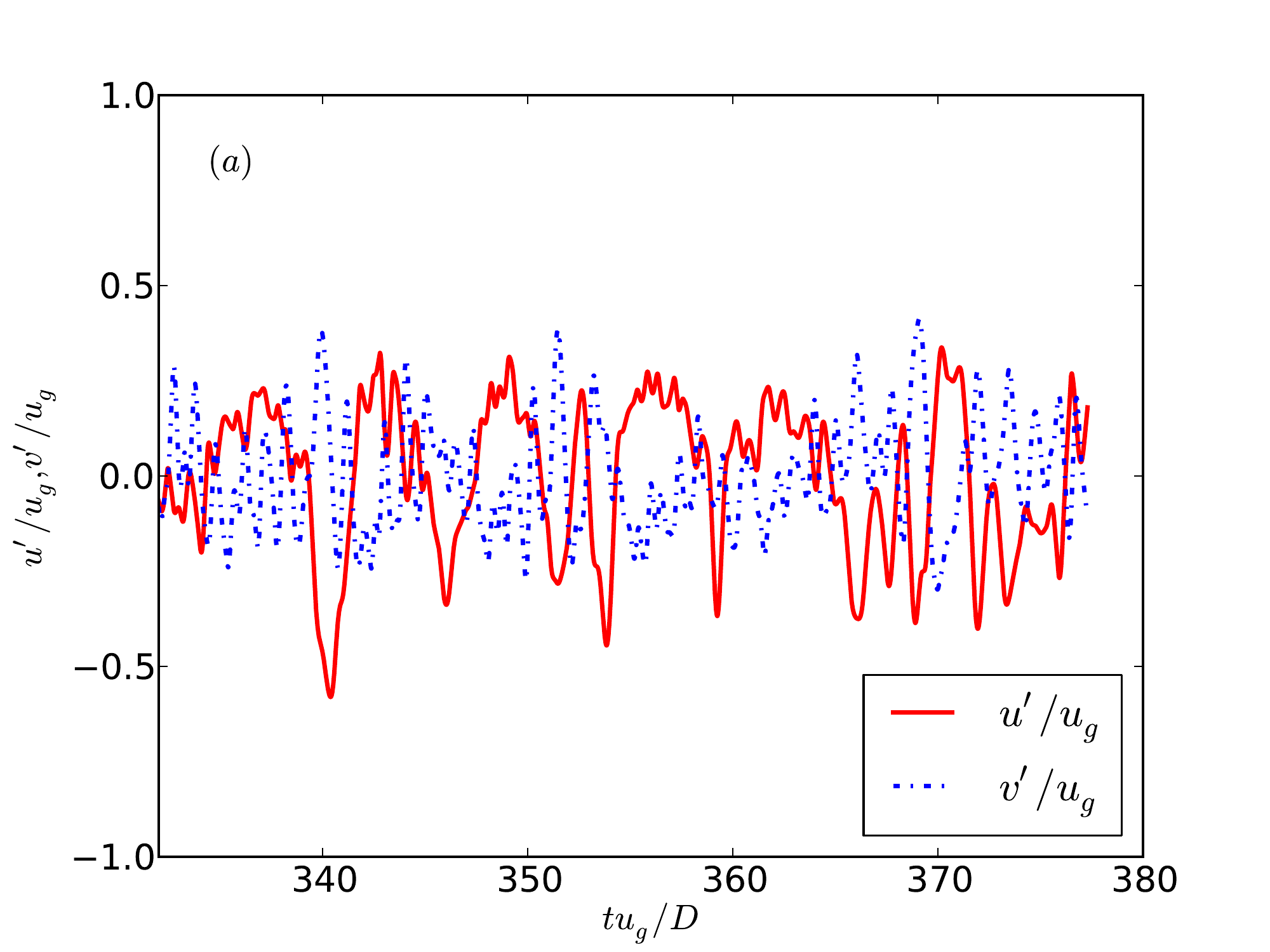}}\\
  \subfloat{\label{fig:Cd0_psd}\includegraphics[trim=0.00in
    0.00in 0.00in 0.00in,clip=true,scale=1.0, width =
    0.9\textwidth]{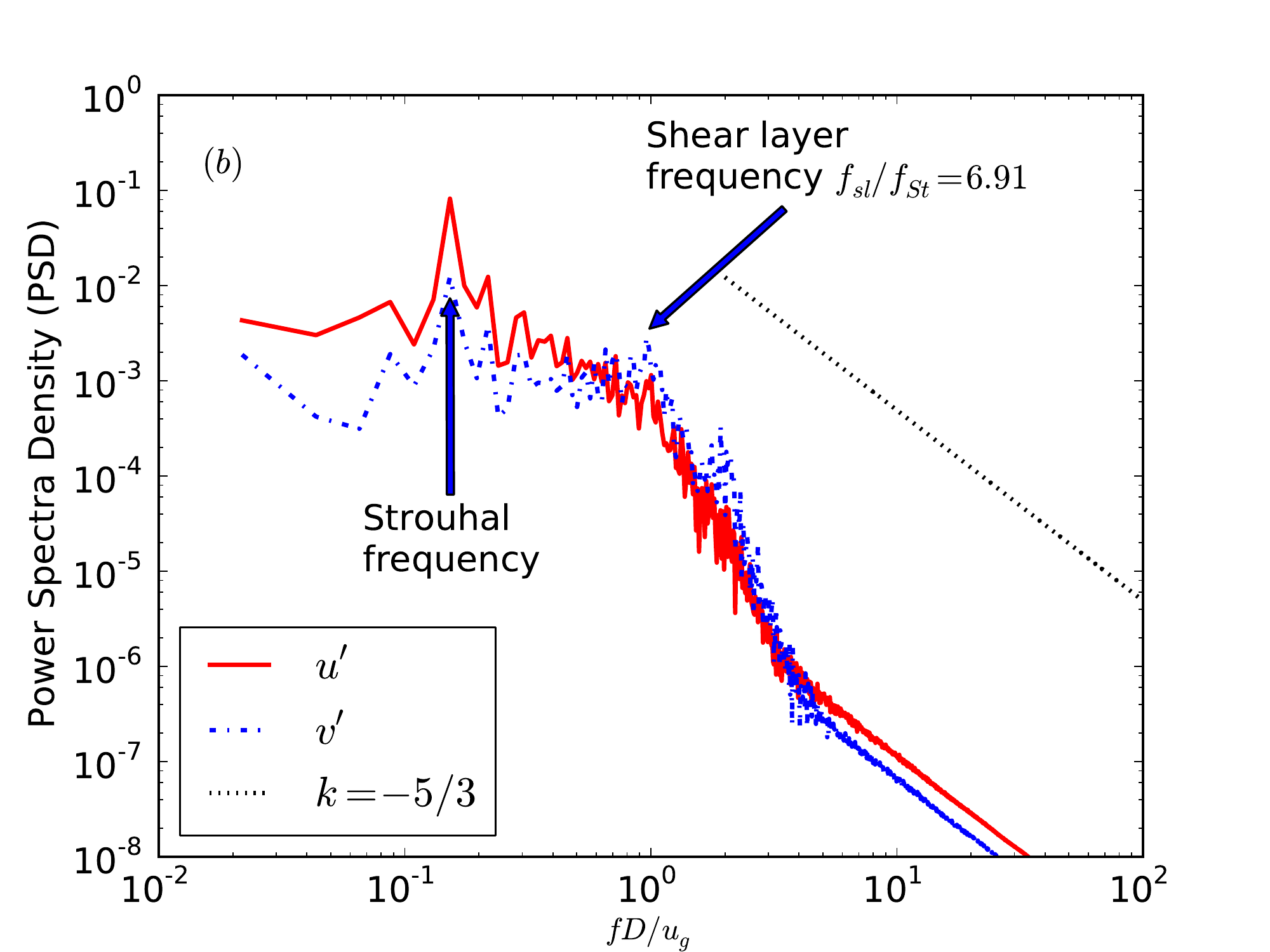}}
  \caption[Time
     histories of velocity signal fluctuations
     behind cylinder \textbf{C1} and the corresponding power spectrum density.]{Time
     histories of velocity signal fluctuations
     behind cylinder \textbf{C1} and the corresponding power spectrum density.}
  \label{fig:Cd0psd}
\end{figure}

Figure~\ref{fig:Cd0psd} presents a statistically significant
sample of time histories of velocity fluctuations at a
point $(x/D = 0.55,y/D = 0.65)$ with respect to the center
of the cylinder that lies in the near wake. The power
spectrum density is obtained by an ensemble average across
the 40 sampling stations in the homogeneous spanwise
direction. The fairly sharp peaks, the Strouhal frequency
$f_{St}$, characterize the predominant vortex
shedding. Moreover, it can be observed that there is another
peak ($f_{sl}$) that represents the frequency of shear-layer
vortexes and is significantly higher and is of a relatively
broader band than $f_{St}$.  For this broadband feature of
shear layer vortexes, \citeauthor{dong2006combined}\cite{dong2006combined} ascribed this
complex phenomena to a few factors, e.g.  the Karman vortex
formation, the varying momentum thickness and the
oscillation of the separation line. However, the value for
$f_{sl}$ predicted for the first column cylinder fails to
match the well-known $Re^{0.67}$ law for an unconfined
circular cylinder (\cite{prasad1997instability}. This
is consistent with the observations from \citeauthor{brun2004role}\cite{brun2004role}
that indicate there is no universal Reynolds number
dependence of $f_{sl}/f_{St}$ for two cylinders placed side
by side.

\begin{figure}[h!tbp]
  \centering
  \subfloat{\label{fig:Cd1_psd}\includegraphics[trim=0.00in
    0.00in 0.00in 0.00in,clip=true,scale=1.0, width =
    0.9\textwidth]{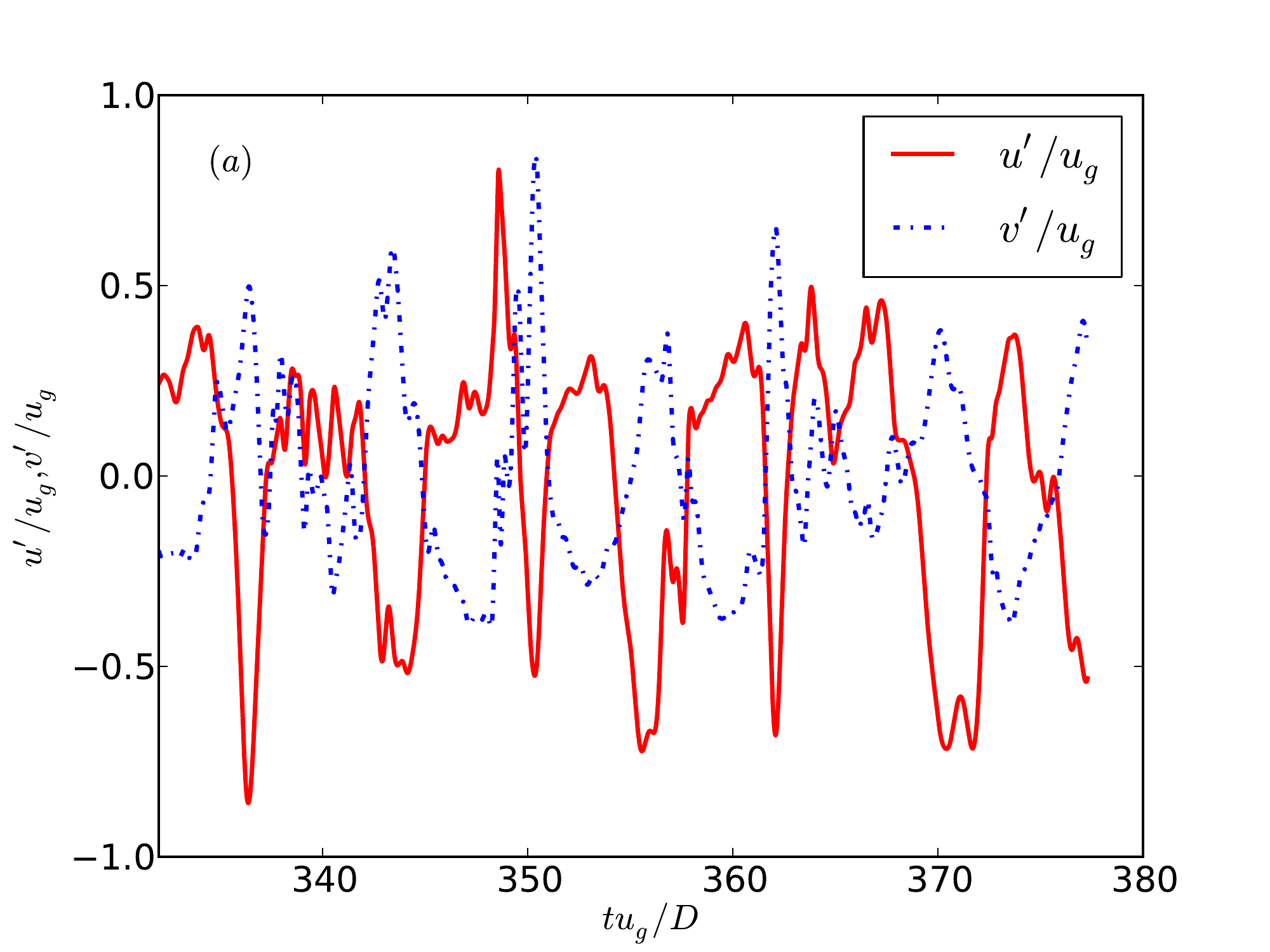}}\\
  \subfloat{\label{fig:Cd1_psd}\includegraphics[trim=0.00in
    0.00in 0.00in 0.00in,clip=true,scale=1.0, width =
    0.9\textwidth]{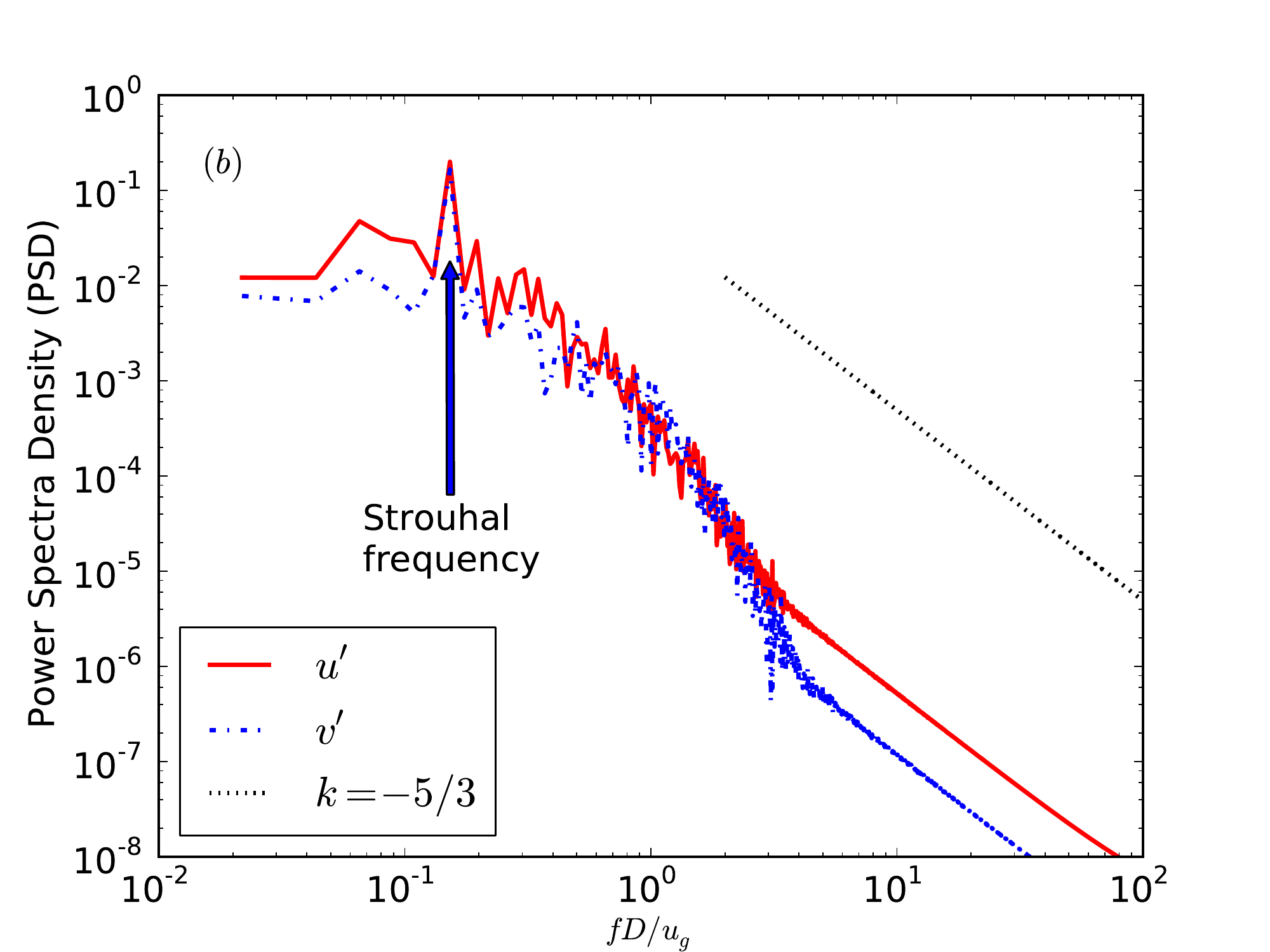}}
  \caption[Time
     histories of velocity signal fluctuations
     behind cylinder \textbf{C2} and the corresponding power spectrum density.]{Time
     histories of velocity signal fluctuations
     behind cylinder \textbf{C2} and the corresponding power spectrum density.}
  \label{fig:Cd1psd}
\end{figure}

\begin{figure}[h!tbp]
  \centering
  \subfloat{\label{fig:Cd3_psd}\includegraphics[trim=0.00in
    0.00in 0.00in 0.00in,clip=true,scale=1.0, width =
    0.9\textwidth]{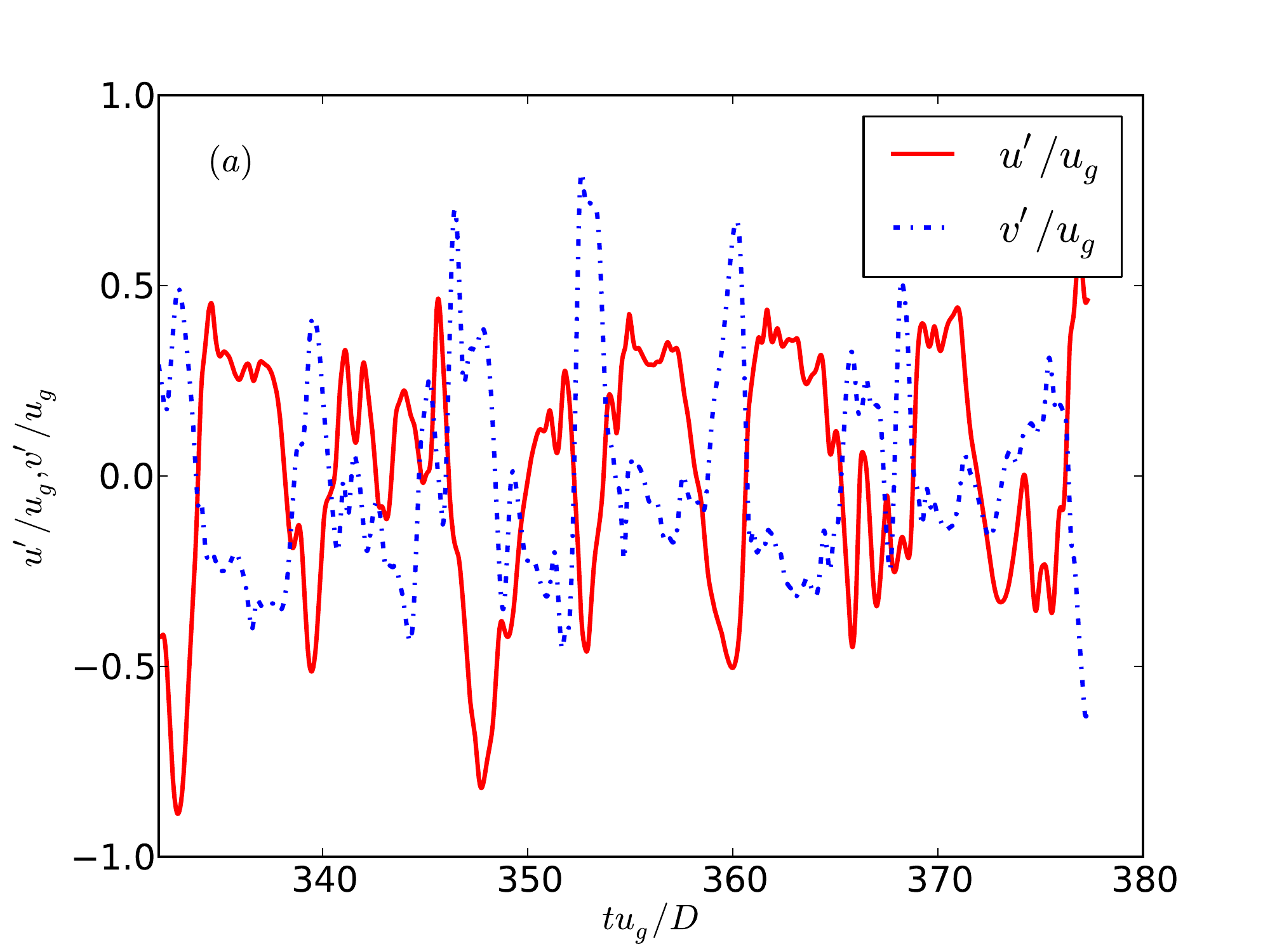}}\\
  \subfloat{\label{fig:Cd3_psd}\includegraphics[trim=0.00in
    0.00in 0.00in 0.00in,clip=true,scale=1.0, width =
    0.9\textwidth]{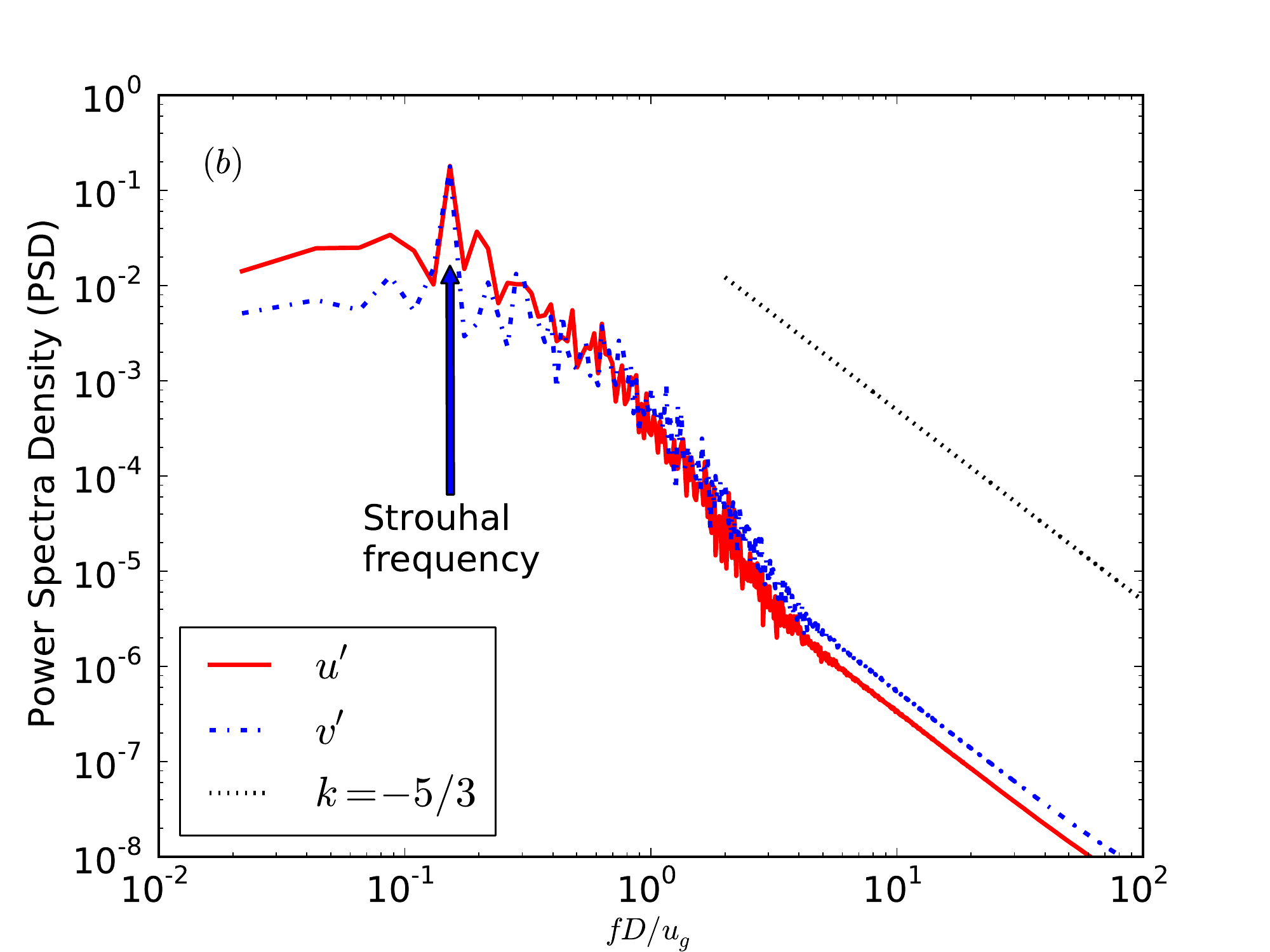}}
  \caption[Time histories of velocity signal fluctuations
  behind cylinder \textbf{C3} and the corresponding power
  spectrum density.]{Time histories of velocity signal
    fluctuations behind cylinder \textbf{C3} and the
    corresponding power spectrum density.}
  \label{fig:Cd2psd}
\end{figure}

\begin{figure}[h!tbp]
  \centering
  \subfloat{\label{fig:Cd4_psd}\includegraphics[trim=0.00in
    0.00in 0.00in 0.00in,clip=true,scale=1.0, width =
    0.9\textwidth]{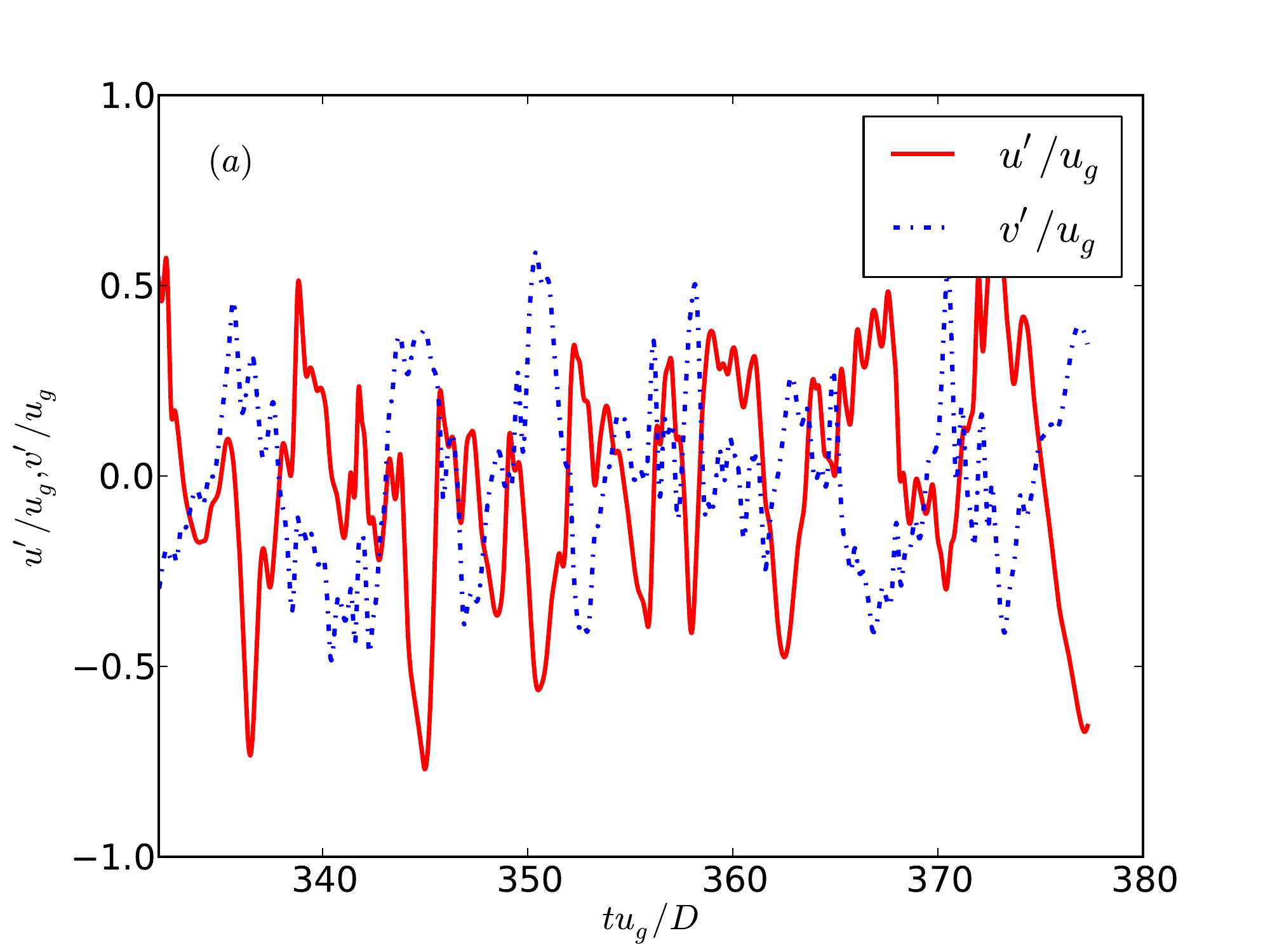}}\\
  \subfloat{\label{fig:Cd4_psd}\includegraphics[trim=0.00in
    0.00in 0.00in 0.00in,clip=true,scale=1.0, width =
    0.9\textwidth]{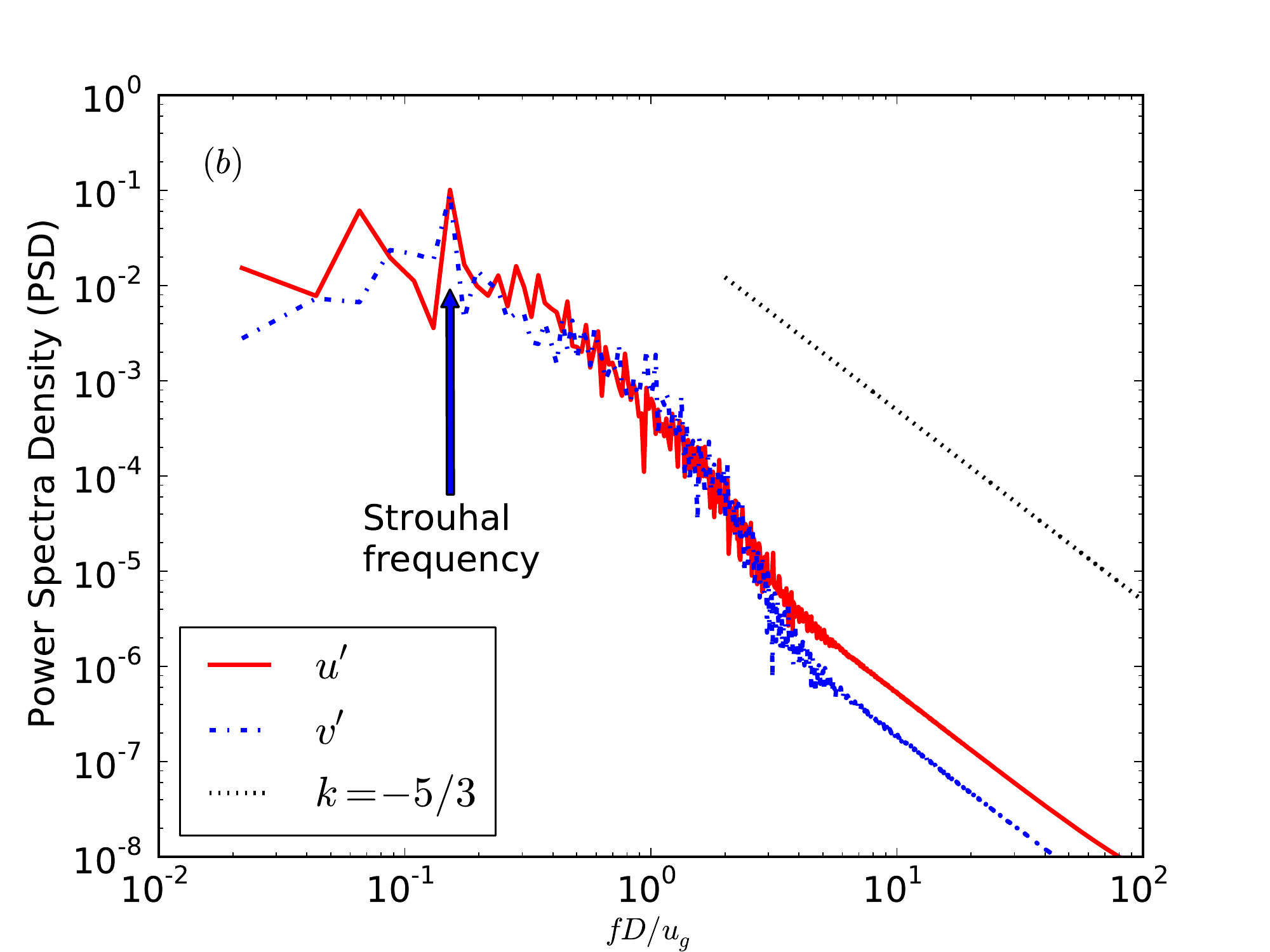}}
  \caption{Time
     histories of velocity signal fluctuations
     behind cylinder \textbf{C4} and the corresponding power spectrum density.}{Time
     histories of velocity signal fluctuations
     behind cylinder \textbf{C4} and the corresponding power spectrum density.}
  \label{fig:Cd3psd}
\end{figure}

The time histories and corresponding power spectrum
densities for the three downstream cylinders are presented
in figure~\ref{fig:Cd1psd}, figure~\ref{fig:Cd2psd} and
figure~\ref{fig:Cd3psd}. The fundamental frequency of vortex
shedding is well pronounced for the three cylinders. From
\citeauthor{gerrard1965disturbance}\cite{gerrard1965disturbance}
and
\citeauthor{gerrard1966mechanics}\cite{gerrard1966mechanics},
the fundamental shedding frequency behaviors like a mean
rather than a fluctuating quantity because the strengths of
the vortexes depend most strongly on the mean rate of
shedding of vorticity, which is governed by the mean
behavior of the separated shear layer. Consequently, it is
reasonable to expect that the fundamental shedding frequency
will show little variations for downstream cylinders in the
sub-critical range of Reynolds number.  Nonetheless, there
is no signature for the shear layer vortexes observed. This
may result from the feature of significant inhomogeneity in
the crosswise direction associated with the wake from the
upstream cylinders.

It is worthwhile noting that the dominant frequency of
vortex shedding predicted in the present study is evaluated
in terms of the mean velocity across the gap $u_g$. It is of
interest to compare the predicted value with the experimental
observations, especially with the universal Strouhal number $St^{*}$
proposed by \citeauthor{roshko1954drag}\cite{roshko1954drag} that is defined
$fd^{*}/u^{*}$ in terms of the wake width between the rows
of vortexes $d^{*}$ and the wake velocity $u^{*}$ obtained
from the free-streamline theory. These are summarized in the
table~\ref{tab:UniSt} along with an extrapolated value from the
measurements for a confined circular cylinder by \citeauthor{richter1976fluctuating}
\cite{richter1976fluctuating}.  It can be observed that
\citeauthor{Shim1985}\cite{Shim1985}'s measurement as well as the present
predicted value fall well within the range $95\%$ of
the universal Strouhal number for the sub-critical range of
Reynolds number.

\begin{table}\small
\centering
\begin{tabular*}{0.65\textwidth}{lcp{3.0cm}}
  \toprule                                                
  
  Case                &               & $St$  \\
  \midrule
  \multirow{4}{*}{Present LES~($Re_g = 15270$)}
                      & C1            & 0.153 \\
                      & C2            & 0.153 \\
                      & C3            & 0.153 \\
                      & C4            & 0.153 \\
  \midrule
  Experiments~\citep{Shim1985}
                      &               & 0.152 \\
  \midrule
  Experiments~\citep{richter1976fluctuating} ($~Re = 1.5 \times 10^{4}$) & $St^{*}$ & 0.151 \\
  \midrule
  Theoretical Value~\citep{roshko1954drag}($~Re < 2 \times 10^5$)      & $St^{*}$ &0.16 \\
\bottomrule
\end{tabular*} 
\caption[Comparison of $St$ with the universal Strouhal number]{Comparison of $St$ with the universal Strouhal number}
\label{tab:UniSt}
\end{table}

\subsection{Correlation length for  vortex shedding}
\begin{sidewaysfigure}
\vspace{6.0in}   
\centering
  \subfloat{\label{fig:C0Corr}\includegraphics[trim=0.00in
    0.00in 0.00in 0.00in,clip=true,scale=1.0, width = 0.4\textwidth]{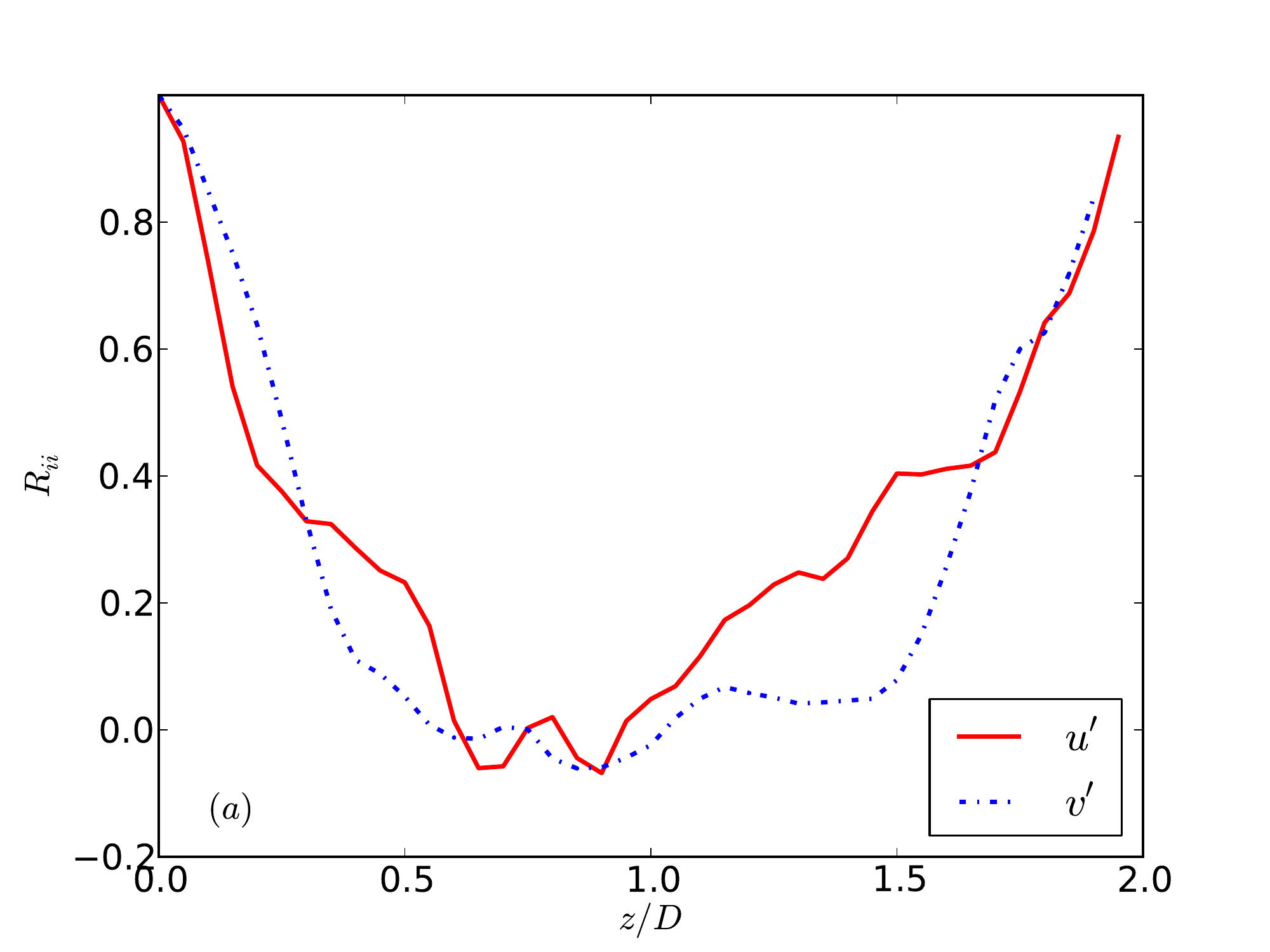}} 
  \subfloat{\label{fig:C1Corr}\includegraphics[trim=0.00in
    0.00in 0.00in 0.00in,clip=true,scale=1.0, width
    =0.4\textwidth]{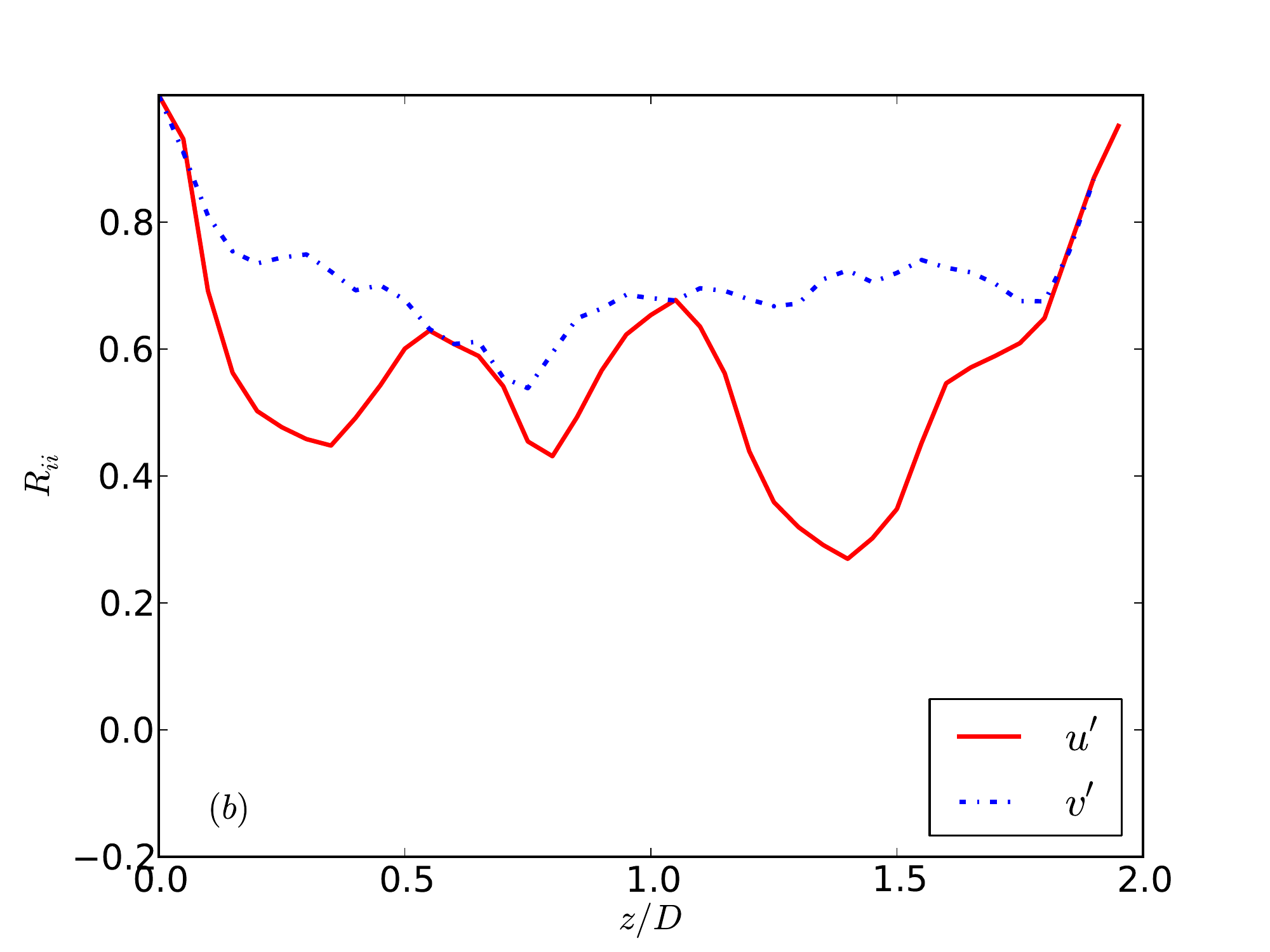}}\\

  \subfloat{\label{fig:C2Corr}\includegraphics[trim=0.00in
    0.00in 0.00in 0.00in,clip=true,scale=1.0, width = 0.4\textwidth]{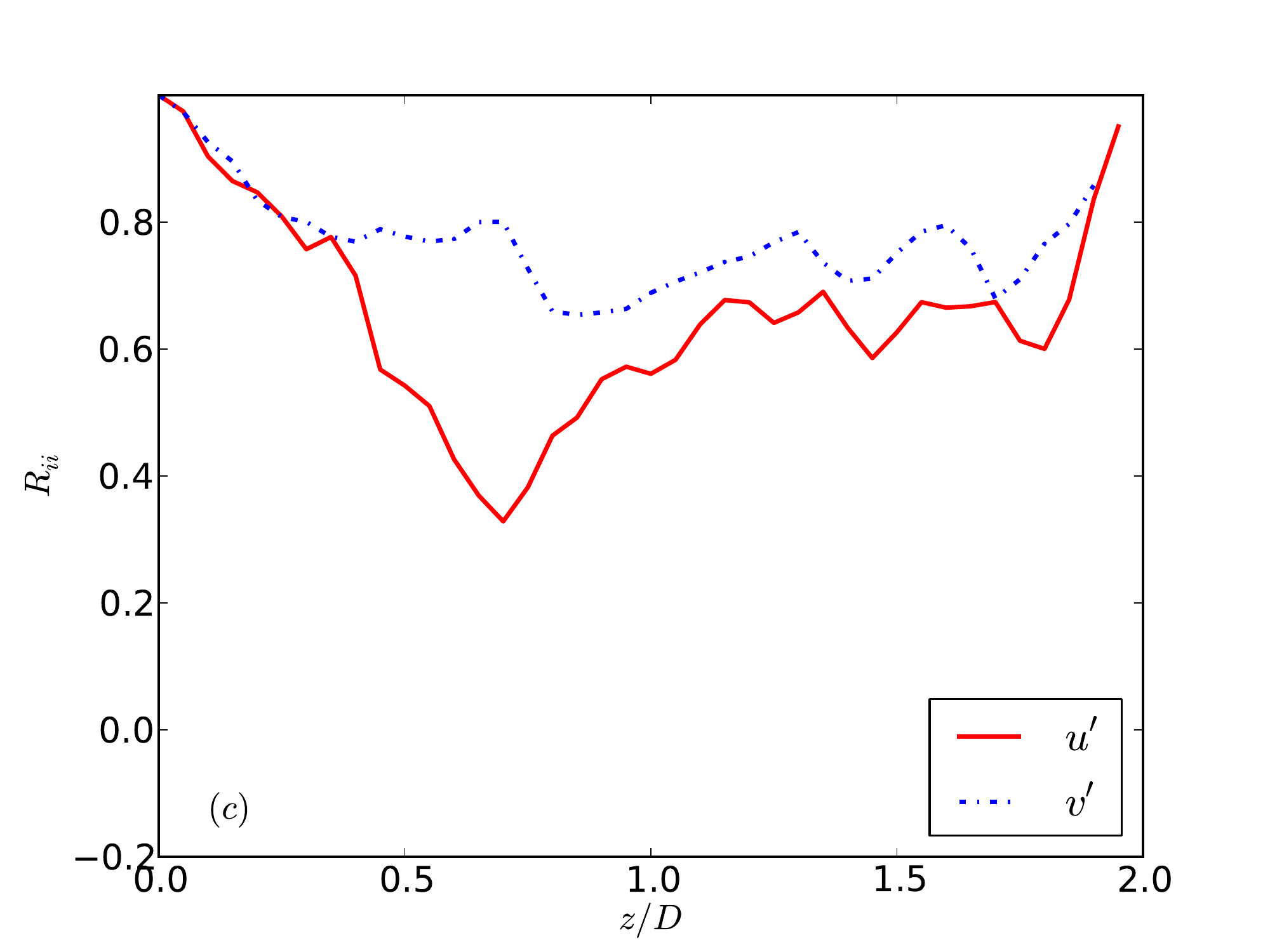}}  
  \subfloat{\label{fig:C3Corr}\includegraphics[trim=0.00in 0.00in
    0.00in 0.00in,clip=true,scale=1.0,width = 0.4\textwidth]{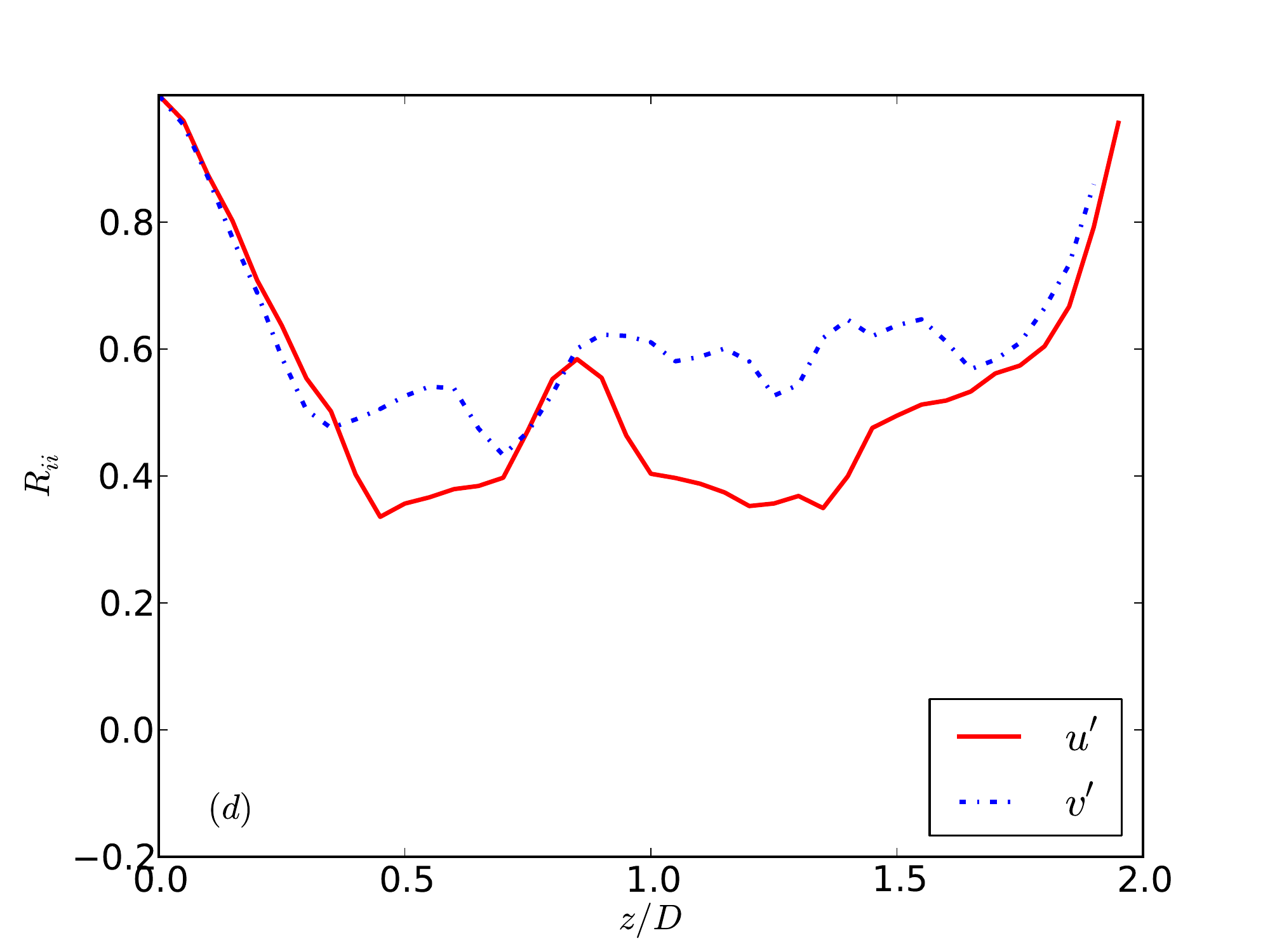}} 
  \caption[Auto-correlation of streamwise and crosswise
  velocity fluctutations as a function of spanwise length
  (a)~\textbf{C1}, (b)~\textbf{C2}, (c)~\textbf{C3}, (c)~\textbf{C4},
  ]{Auto-correlation of streamwise and crosswise
  velocity fluctutations as a function of spanwise length
  (a)~\textbf{C1}, (b)~\textbf{C2}, (c)~\textbf{C3}, (c)~\textbf{C4},
  }
  \label{fig:Correlation}
\end{sidewaysfigure}

To examine the spatial structure of vortex shedding behind
the cylinders, figure~\ref{fig:Correlation} presents the
auto-correlation functions for the streamwise and crosswise
velocity components for the 40-sampling stations $(x/D =
0.55,y/D = 0.65)$ with respect to the axis of the cylinder
the across the homogeneous spanwise direction. It is defined
as
\beq 
R_{ii}(\mathbf{x};\mathbf{z}, t) =
\frac{\overline{u_{i}^{\prime}(\mathbf{x}; t)u_{i}^{\prime}(\mathbf{x+z};
    t)}}{\overline{{u_{i}^{\prime}}^2(\mathbf{x}; t)}}.
\eeq
From figure~\ref{fig:C0Corr}, it can be observed for the
first column cylinder C1 that $R_{u^{\prime}u^{\prime}}$ and
$R_{v^{\prime}v^{\prime}}$ are decreasing monotonically to
zero within the range of $L/D = 1$. This implies that the
spanwise length of the biggest eddy from vortex shedding
approximately equals to the diameter of the cylinder. This
feature has been demonstrated for an unconfined single
circular cylinder by previous researchers. Nonetheless, the
fact that $R_{u^{\prime}u^{\prime}}$ and
$R_{v^{\prime}v^{\prime}}$ do not tend asymptotically
towards zero at large separation distance is concerned with
the periodic boundary condition employed for the homogeneous
spanwise direction in the present LES study.

Nevertheless, from
figure~\ref{fig:C1Corr},\subref*{fig:C2Corr}
\subref*{fig:C3Corr}, it can be observed that the downstream
cylinders C1, C2 and C3 display distinct behaviors with
respect to the auto-correlation as a function of the
spanwise length for the streamwise and crosswise velocity
fluctuations. First, within the length of $L/D = 1$,
$R_{u^{\prime}u^{\prime}}$ and $R_{v^{\prime}v^{\prime}}$ do
not decease to zero. Secondly, $R_{v^{\prime}v^{\prime}}$
wiggles across the middle part of the cylinder $L/D \in
\left[0.5, 1.5\right]$. For both of the discrepancies from
the first column cylinder C1, it may result from the mixing
of the shedding vortexes from different column cylinders,
hence it gives rise to complex eddy patterns of the wake.   

\section{Concluding remarks}
Turbulent flow across in-line tube-banks with transverse and
longitudinal pitch $P_T/D = 2.67$ and $P_L/D = 2.31$,
respectively, has been studied successfully by Large Eddy
Simulation (LES) based on the dynamic Smagorinsky subgrid
scale model (SGS) with a wall-layer model. Flow structures
across the tube-banks based on the normalized Q criterion is
presented. The middle cylinder from each column is chosen to
present results and compared with experiments. The surface
pressure characteristics observed in \citeauthor{Shim1985}\cite{Shim1985}'s
experiment are well reproduced irrespective of some
discrepancies that can be attributed to the difficulty in
numerically mimicing the inflow condition of the experiment.
Quite satisfying agreement is observed between the
simulation and experimental observations for the drag and lift
coefficients, which indicates the second column cylinder
experiences the minimum drag force and maximum lift force
fluctuation. A frequency analysis for velocity signals at
the position with respect to each cylinder axis ($(x/D = 0.55,y/D = 0.65)$) is
presented and compared with experimental as well as
theoretical work. These results show that the dominant
vortex shedding frequency does not show variations across
the tube bank. Nevertheless, the instability
frequency of shear layer is not observed for the downstream cylinders. As
far as the first column cylinder is concerned, the shear
layer instability observed does not show agreement with the universal value
for an unconfined single circular cylinder; however, this
supports the recent experimental measurements by
\citeauthor{brun2004role}\cite{brun2004role}. Finally, auto-correlation functions for
streamwise and cross velocity fluctuations as a function of
the spanwise length are investigated. They indicate the
turbulent eddy behind downstream cylinders are of more
complex structure than the first column cylinder as result
of the mixing shedding vortexes from different column
cylinders.

\section{Acknowledgments}
We wish to acknowledge the support of British Energy (Part
of EDF).

\bibliographystyle{model1-num-names}
\bibliography{paper3_FTaC_2016_r0_arXiv}

\pagebreak
\clearpage

\clearpage

\clearpage

\clearpage

\clearpage

\clearpage

\clearpage

\clearpage
\clearpage

\clearpage

\clearpage

\clearpage

\clearpage

\clearpage

\end{document}